\documentclass[prd,aps,preprint,onecolumn,superscriptaddress,nofootinbib,eqsecnum,preprintnumbers, longbibliography]{revtex4-1}

\pdfoutput=1

\usepackage{amsmath, latexsym, amssymb, graphicx, ifpdf, slashed, color, hyperref, url, cancel, bbm}

\usepackage[T1]{fontenc}

\usepackage[usenames,dvipsnames]{xcolor}

\hypersetup{colorlinks, citecolor=nicegreen,linkcolor=nicered, urlcolor=niceblue}
\definecolor{nicered}{rgb}{.7,.1,.1}
\definecolor{nicegreen}{rgb}{.2,.7,.1}
\definecolor{niceblue}{rgb}{0.1,0.2,0.6}
\definecolor{darkblue}{rgb}{0,0,.5}

\begin{document}


\title{\texorpdfstring{
Anatomy of Diluted Dark Matter in the\\ Minimal Left-Right Symmetric Model}{
Anatomy of Diluted Dark Matter in the Minimal Left-Right Symmetric Model}}

\author{Miha Nemev\v{s}ek}
\email{miha.nemevsek@ijs.si}
\affiliation{Jo\v{z}ef Stefan Institute, Jamova 39, 1000 Ljubljana, Slovenia}
\affiliation{Faculty of Mathematics and Physics, University of Ljubljana, Jadranska 19, 
1000 Ljubljana, Slovenia}

\author{Yue Zhang\,}
\email{yzhang@physics.carleton.ca}
\affiliation{Department of Physics, Carleton University, Ottawa, ON K1S 5B6, Canada}

\date{\today}

\vspace{1cm}

\begin{abstract} 
Temporary matter domination and late entropy dilution, injected by a ``long-lived'' particle in the early universe, serves as a standard mechanism for yielding the correct dark matter relic density. We recently pointed out the cosmological significance of diluting particle's partial decay into dark matter. When repopulated in such a way, dark matter carries higher momentum than its thermal counterpart, resulting in a suppression of the linear matter power spectrum that is constrained by the large scale structure observations. In this work, we study the impact of such constraints on the minimal left-right symmetric model that accounts for the origin of neutrino mass. We map out a systematic anatomy of possible dilution scenarios with viable parameter spaces, allowed by cosmology and various astrophysical and terrestrial constraints. We show that to accommodate the observed dark matter relic abundance the spontaneous left-right symmetry breaking scale must be above PeV and cosmology will continue to provide the most sensitive probes of it. In case the dilutor is one of the heavier right-handed neutrinos, it can be much lighter and lie near the electroweak scale.
\end{abstract}

\preprint{}

\maketitle
\tableofcontents

%
%
\section{Introduction} \label{sec:Intro}

The field of particle physics and cosmology is facing at least three unresolved issues, driven by
experiments: the nature of dark matter, the origin of neutrino mass, and the origin of the 
matter-anti-matter asymmetry in the universe.
They are likely linked to new fundamental laws of nature. 
Conceptually, it would be very appealing to have these puzzles solved within a single unified 
framework.

A heavy neutrino is one of the oldest, simplest and most obvious of dark matter candidates. 
It was first introduced as a Standard Model (SM) gauge singlet with a small mixing with the 
active neutrinos, produced via active-sterile neutrino oscillations in the early universe.
With improved astrophysical observations, both the Dodelson-Widrow~\cite{Dodelson:1993je} and 
the Shi-Fuller~\cite{Shi:1998km} mechanisms are already excluded~\cite{DES:2020fxi, Abazajian:2017tcc}.
To save such oscillation mechanisms, one must resort to novel neutrino 
self-interactions~\cite{DeGouvea:2019wpf, Kelly:2020pcy, Kelly:2020aks, An:2023mkf}. 
A common assumption here is a va\-nish\-ing dark matter population at very early times, which can 
easily be affected by high-scale new physics. 
Right-handed neutrinos are often mandatory for gauge anomaly cancellation in many extensions of 
the SM~\cite{Pati:1974yy, Fritzsch:1974nn, Mohapatra:1979ia}.
Assuming the universe was once sufficiently hot, new gauge interactions can then bring them into 
thermal equilibrium with the SM. 
A right-handed neutrino can be made cosmologically stable and comprise 100\% of dark matter in 
the universe. 
While it appears nearly sterile at low energies, the origin of dark matter (its abundance
and momentum distribution) is governed by details of the high-scale theory.

The minimal left-right symmetric model (LRSM), based on $SU(3)_c\times SU(2)_L\times SU(2)_R \times 
U(1)_{B-L}$~\cite{Mohapatra:1974gc, Senjanovic:1975rk, Senjanovic:1978ev}, was originally proposed 
as a theory for non-zero neutrino mass~\cite{Mohapatra:1979ia}.
In the model, parity is implemented as a left-right $Z_2$ symmetry, acting between left- and
right-handed fermions.
Since $SU(2)_R$ is gauged, three generations of right-handed neutrinos need to be present 
to cancel the anomalies.
Parity and new gauge symmetries are spontaneously broken above the electroweak scale and 
light neutrino masses originate from both type-I and II seesaw~\cite{Mohapatra:1979ia}.

Remarkably enough, the Dirac couplings also get predicted~\cite{Nemevsek:2012iq}, which makes
the LRSM a particularly complete and predictive theory for neutrino masses. 
The breaking of lepton number manifests itself in a number of processes, ranging from direct 
production at high energy colliders~\cite{Keung:1983uu, Nemevsek:2011hz}, neutrino-less double
beta decay~\cite{Mohapatra:1980yp, Tello:2010am}, LFV~\cite{Cirigliano:2004mv, Nemevsek:2011aa} 
and cosmology~\cite{Mohapatra:1992pk, Joshipura:2001ya}. 
Moreover, the model can explain the origin of cosmic baryon asymmetry via leptogenesis, as long as
the LR scale is sufficiently high~\cite{Frere:2008ct, BhupalDev:2014hro}.

The lightest right-handed neutrino (called $N_1$ hereafter) as a viable dark matter candidate 
in the minimal LRSM was first considered by Bezrukov,  Hettmansperger and Lindner~\cite{Bezrukov:2009th}. 
It features a dark matter candidate with a light mass below $\sim$\,MeV, in order to be 
cosmologically stable. 
On the other hand, its mass is bounded from below by several keV scale from phase space packing 
in dwarf galaxies, because $N_1$ is a fermion and subject to Pauli blocking~\cite{Tremaine:1979we, Boyarsky:2008ju}. 
Slightly stronger mass lower bound applies if $N_1$ had thermal contact with the SM plasma in the early universe (warm dark matter)~\cite{Viel:2013fqw, Irsic:2017ixq, Gilman:2019nap, Hsueh:2019ynk, DES:2020fxi, Enzi:2020ieg, Nadler:2021dft}.
The right-handed current gauge interaction, mediated by the $W_R$, decouples in the 
early universe and always leaves $N_1$ freezing out ultra relativistically. 
This overproduces the dark matter relic abundance, unless there is a ``long-lived'' matter component 
that temporarily dominates the energy density of the universe before decaying dominantly into the 
SM~\cite{Scherrer:1984fd}.
The late entropy release causes a relative dilution of the final dark matter abundance and brings it
down to the observed value. 
In~\cite{Bezrukov:2009th}, it was suggested that the role of the diluting particle can be taken on
within the LRSM by one of the heavier right-handed neutrinos. 
The corresponding $W_R$ boson mass scale is typically constrained to be rather high to facilitate 
the relativistic freeze-out and longevity of the diluting particle. 
The possibility of having a lighter $W_R$ boson was investigated in~\cite{Nemevsek:2012cd},
which resorts to a decaying phase space suppression to keep the lifetime 
of the diluting particle sufficiently long.

Recently, the dark matter dilution mechanism was revisited and a new, model-in\-de\-pen\-dent constraint 
has been discovered~\cite{Nemevsek:2022anh}.
This new opportunity for testing dark matter dynamics lies in the partial decay of the dilutor into 
dark matter. 
Such decay modes exist quite generically, either at tree or loop level, and the branching ratio is 
sometimes fixed by the internal structure of a UV complete model.
With dark matter re-populated this way, the relic density obtains a secondary component 
on top of the primary one that comes from the usual freeze-out.
Most importantly, this component is predicted to be much more energetic than the original thermal one.

Under a reasonable assumption that both the dark matter and the dilutor freeze-out 
relativistically~\footnote{We will comment on what happens to the constraint from dilution 
when this assumption is relaxed. 
In the LRSM, relativistic freeze-out is always valid, regardless of which particle plays the
role of the dilutor.
}, 
the secondary dark matter particles stay relativistic until the temperature of the 
universe cools down to around eV scale. 
This temperature is nearly independent of parameters including the dark matter mass and the dilutor's mass and lifetime.
As a result, dark matter free-streaming strongly impacts the matter power spectrum and the formation of 
large scale structures.
Using the existing data from the Sloan Digital Sky Survey (SDSS), an upper bound on the branching 
ratio for dilutor into dark matter is set at about $\lesssim 1\%$. 
Because the primordial perturbations remain linear on large scales, a robust cosmological constraint 
can be applied on the fundamental theories for the origin of dark matter.

In the context of LRSM, the decay of a heavy right-handed neutrino to the lighter one could occur via 
the exchange of the $W_R$ gauge boson, similar to weak decays in the SM. 
If this is the dominant mode, the branching ratio is predicted by the number of lighter fermions and
universality of the $SU(2)_R$ gauge interactions and comes out to be larger than than 10\%, which turns out
to be forbidden by the large scale structure observations~\cite{Nemevsek:2022anh}.
To mitigate this exclusion, the right-handed neutrino dilutor must have other significant decay channels
to reduce the dark matter re-population.
Such an important constraint has been ignored in previous analysis. 
We are therefore strongly motivated to revisit the viability of right-handed neutrino dark matter in 
the LRSM and, as we shall see, they strongly affect the allowed parameter space where an appropriate 
dark matter relic density can be obtained. 
In performing a systematic and thorough analysis, we first focus on the usual dilutor in the form
of another right-handed neutrino, and then identify a new candidate from the scalar sector of the LRSM
that can also play the role of dilution.

Before our journey begins, we would like to stress that the entropy dilution explored here in the context 
of LRSM is a generic new physics scenario for fixing the dark matter relic density, or for suppressing the amount of extra radiation ($\Delta N_{\rm eff}$) in the early universe. It has been  employed in 
a broad range of dark matter models including the gravitino, twin-Higgs models, and various dark 
sector incarnations~\cite{Moroi:1999zb, Baltz:2001rq, Asaka:2006ek, Hasenkamp:2010if, Arcadi:2011ev, 
Zhang:2015era, Patwardhan:2015kga, Chacko:2016hvu, Cirelli:2016rnw, Soni:2017nlm, Contino:2018crt, Evans:2019jcs, Cosme:2020mck, Dror:2020jzy, 
Chanda:2021tzi, Asadi:2021bxp, Baumholzer:2021heu, Bleau:2023fsj}. 
The same large scale structure constraint would also affect the viability of dark matter in these models 
and must be taken into account in future studies.

This article is organized as follows. 
In section~\ref{sec:LRSM}, we give a lightning review of the minimal LRSM and highlight several aspects 
of the model that are important for the dark matter study in this work. 
In section~\ref{sec:DMBasics}, we discuss the entropy dilution mechanism that features temporary matter
domination by a ``long-lived'' diluting particle. 
We go beyond the earlier work~\cite{Nemevsek:2022anh} and provide a detailed derivation of the Boltzmann
equation and distribution function of secondary dark matter component from dilutor's decay.
The discussion in this section is model-independent and easily applicable to other models, besides
the LRSM, that resort to a similar dilution mechanism. 
In section~\ref{sec:theAnatomy}, we present the anatomy of right-handed neutrino dark matter in the minimal LRSM 
by exhausting all the possible dilution scenarios that we have envisioned.
This includes the heavier right-handed neutrino or the Higgs boson counterpart of left-right symmetry 
breaking (the right-handed triplet) playing the role of dilutor. 
For each of the cases, we establish the viable parameter space for DM, compatible with the large scale 
structure limits from SDSS, along with other constraints that include the correct relic density, 
generation of neutrino mass, big-bang nucleosynthesis and $X$-ray line searches.
We also comment on supernova cooling and existing laboratory constraints on the LR scale. 
We conclude and provide an outlook of opportunities in section~\ref{sec:conclusion}.

%
%
\section{Dark matter in the minimal left-right symmetric model} 
\label{sec:LRSM}

We start with a brief overview of the structure of the minimal LRSM and highlight several ingredients 
that are important for understanding the cosmology in the later sections.
We refer to~\cite{Mohapatra:1974gc, Senjanovic:1975rk, Senjanovic:1978ev} for the original works 
and in-depth reviews of the model.

%
\subsection{The minimal left-right symmetric model}

The LRSM is based on the gauge group $G_{\text{LR}} = SU(3)_c\times SU(2)_L \times SU(2)_R \times 
U(1)_{B-L}$, with a discrete $Z_2$ symmetry interchanging the left and right $SU(2)$ sectors.
At low energies, the $Z_2$ symmetry manifests itself as the parity symmetry of QCD and QED.
Quarks and leptons come in parity symmetric representations
\begin{equation}\label{eq:FermionReps}
\begin{split}
	Q_L &= \begin{pmatrix} u_L \\ d_L \end{pmatrix}    = \left(3, 2, 1, \frac{1}{3} \right) \, , \quad
    Q_R  = \begin{pmatrix} u_R \\ d_R \end{pmatrix}    = \left(3, 1, 2, \frac{1}{3} \right) \, , 
    \\
	L_L &= \begin{pmatrix} \nu \\ \ell_L \end{pmatrix} = \left(1, 2, 1, -1 \right) \, , \quad
    L_R  = \begin{pmatrix} N   \\ \ell_R \end{pmatrix} = \left(1, 2, 1, -1 \right) \, ,
\end{split}
\end{equation}
where $N$ stands for the right-handed neutrino.
The scalar potential of the minimal model is also parity symmetric.
It consists of three complex fields: a bi-doublet $\Phi = \left( 1, 2, 2, 0 \right)$ and two triplets 
$\Delta_L = \left( 1, 3, 1, 2 \right)$ and $\Delta_R = \left( 1, 1, 3, 2 \right)$ under $G_{\text{LR}}$,
with the following field assignments
\begin{align} \label{eq:ScalarReps}
  \Phi &= 
  \begin{pmatrix} \phi_1^0 & \phi_2^ + 
  \\
  \phi_1^- & \phi_2^0 \end{pmatrix}  \, , 
  &
  \Delta_{L, R} &= 
  \begin{pmatrix} \delta^+ /\sqrt{2} & \delta^{++} 
  \\
  \delta^0 & -\delta^{+}/\sqrt{2}
  \end{pmatrix}_{L, R}  \, .
\end{align}
Under parity, $\Phi\to \Phi^\dagger$, $\Delta_L \leftrightarrow \Delta_R$.
Starting from the LR and parity symmetric potential, it was shown that parity is broken 
spontaneously~\cite{Senjanovic:1975rk, Senjanovic:1978ev}, with either with two doublets~\cite{Senjanovic:1978ev} 
or two triplets~\cite{Mohapatra:1979ia, Mohapatra:1980yp}.
The complete form of the potential was discussed in~\cite{Basecq:1986rw} and studied in some depth
in subsequent years~\cite{Gunion:1989in, Kiers:2002cz, Deshpande:1990ip, Duka:1999uc, Khasanov:2001tu}
with more recent works focusing on phenomenological signals~\cite{Dekens:2014ina, Bambhaniya:2015wna, Dev:2016dja}.
A strong lower bound from perturbativity of the potential was worked out in~\cite{Maiezza:2016bzp}, with constraints
from vacuum the vacuum structure~\cite{BhupalDev:2018xya} and opportunities for gravitational waves~\cite{Brdar:2019fur}.

The above quantum numbers allow for the following Yukawa terms that couple the fermions to scalars
\begin{equation}\label{eq:Yukawa}
\begin{split}
	\mathcal{L}_{\text{Yuk}} = 
    & \bar Q_L \left( Y_q \, \Phi + \tilde Y_q \, \tilde \Phi \right) Q_R + 
    \bar L_L \left( Y_l \, \Phi + \tilde Y_l \, \tilde \Phi \right) L_R
    \\
	& + 
    Y_{\Delta_L} L_L^T i \sigma_2 \, \Delta_L \, L_L + 
    Y_{\Delta_R} L_R^T i \sigma_2 \, \Delta_R \, L_R + \text{h.c.} \, .
\end{split}
\end{equation}
where $\tilde \Phi = i\sigma_2 \Phi^* i\sigma_2$ and we suppressed the family indices.
The $\sigma_2$ matrices operate within the two $SU(2)_{L, R}$ group spaces and ensure
gauge invariance.
The first two terms are of the Dirac type and give the usual mass terms that connects the left
and right chiral fields, as in the SM.
A new feature is the Dirac term for neutrinos and the Majorana-type Yukawa couplings $Y_{\Delta_{L,R}}$
in the second line, that generate lepton number violating masses for neutrinos.

The spontaneous symmetry breaking occurs in two steps.
First, the $SU(2)_R\times U(1)_{B-L}$ symmetry is broken down to $U(1)_Y$ for hypercharge by
the vacuum expectation value (VEV) of the right-handed scalar triplet 
$\langle \Delta_R^0 \rangle = v_R/\sqrt{2}$, which lies well above the electroweak scale. 
It generates masses for the new gauge bosons $W_R^\pm$ and $Z'$, with a mass relation $M_{Z'}=\sqrt{3} M_{W_R}$. 
Through the $Y_{\Delta_R}$ Yukawa coupling term in Eq.~\eqref{eq:Yukawa}, the $v_R$ condensate also 
gives a Majorana mass to the right-handed neutrinos $N$.

The second stage of spontaneous symmetry breaking is triggered by the VEV of the bi-doublet scalar,
$\langle \phi_1^0\rangle = v\cos\beta/\sqrt{2}$ and $\langle \phi_2^0\rangle = v\sin\beta e^{i\alpha}/\sqrt{2}$, 
where $v = 246 \, $GeV.
Without loss of generality, we choose $\beta \in (0, \pi/2)$.
In view of $SU(2)_L$, the bi-doublet effectively behaves as two Higgs doublets, thus their VEVs 
can give masses to the regular $W^\pm$ and $Z$ gauge bosons that mediate the weak interactions.
Through the Yukawa coupling in Eq.~\eqref{eq:Yukawa}, the electroweak VEVs also provide Dirac 
mass matrices for all the fermions, including the one between left-handed neutrinos $\nu$ and 
right-handed ones $N$. 

%
%
\subsection{Essential model ingredients for the early universe}

\subsubsection{Right-handed currents}

The most important ingredient of the LRSM, relevant for dark matter cosmology, are the new gauge 
interactions, mediated by the $W_R^\pm$ and $Z'$ gauge bosons. 
The right-handed charged-current interactions for fermions take on the form
\begin{equation} \label{eq:RHCurrent}
 {\mathcal L}_{\text{gauge}} = 
 \frac{g}{\sqrt{2}} W_R^\mu \left( 
    \bar{N}   \gamma_\mu {V_{\text{PMNS}}^{R\dag}} \ell_R + 
    \bar{u}_R \gamma_\mu {V_{\text{CKM} }^R}       d_R   \right) +\text{h.c.} \, , 
\end{equation}
where all the fermions fields are now in their mass eigenstates. 
We have introduced the right-handed CKM and PMNS matrices that transform the fermions from the
weak flavor into the mass basis. 
Here we neglected the mixing between the $W$ and $W_R$ gauge bosons, which is constrained to be small (see Sec.~\ref{sec:WWRmix}) and its effects in dark matter pehenomenology will be accounted for in \ref{sec:type2Xray} and \ref{appdx1}.

In the minimal LRSM with parity symmetry, the right-handed CKM matrix has been solved both 
numerically~\cite{Zhang:2007da, Zhang:2007fn, Maiezza:2010ic} and 
analytically~\cite{Senjanovic:2014pva, Senjanovic:2015yea}.
Its off-diagonal elements are suppressed by similar powers of the Wolfenstein parameter as the 
regular CKM matrix, but all the matrix elements carry extra phase factors with implications for the
strong CP violation~\cite{Maiezza:2010ic, Maiezza:2014ala, Bertolini:2019out, Bertolini:2020hjc}.  
For the purposes of this work, the exact form of the quark charged currents is not really important
and we shall approximate $V_{\rm CKM}^R$ with a unit matrix.
The situation is quite different for the leptons.
There, the form of right-handed PMNS matrix is much less constrained and allowed to be wildly different 
from the regular PMNS matrix.
The only special case is when the light neutrino mass contribution is dominated by the type-II 
seesaw contribution. 
In this case, parity requires that $V_{\rm PMNS}^R$ be identical to its left-handed counterpart.

\subsubsection{Lightest right-handed neutrino as dark matter}
\label{sec:N1DM}

\begin{figure}[!ht]
  \begin{center}
    \includegraphics[width=0.45\textwidth]{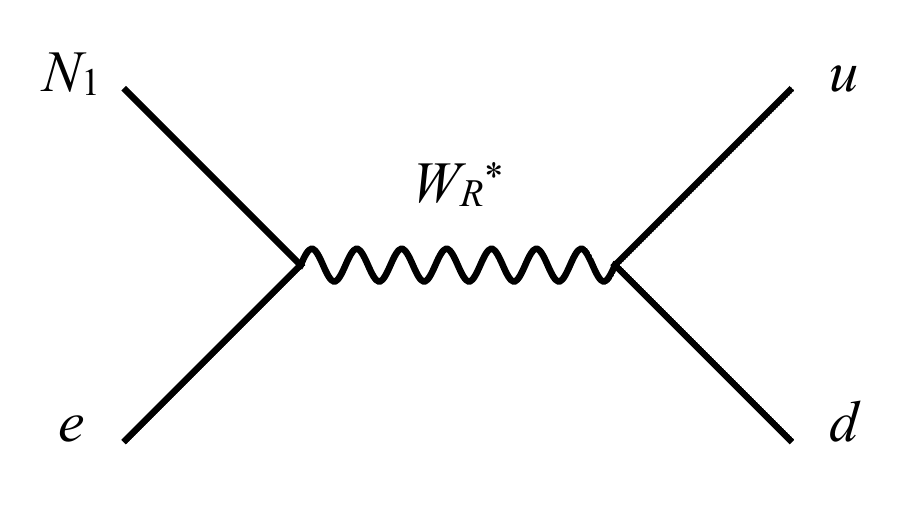}\vspace{-0.7cm}
  \end{center}
  \caption{Feynman diagram for right-handed charged-current interaction of dark matter $N_1$ in LRSM.}
  \label{fig:FeynDiagN1}
\end{figure}

The minimal LRSM does not preserve any exact $Z_2$ symmetry that would stabilize DM, even LR parity 
gets broken spontaneously and its quality is not crucial for dark matter stability.
As a result, none of the new particles beyond the SM are absolutely stable. 
As argued in~\cite{Nemevsek:2012cd}, the only candidate for dark matter in the model is the lightest 
right-handed neutrino, $N_1$~\footnote{\label{footVacuum}The real part of $\Delta_R^0$ in~\eqref{eq:ScalarReps} may also be
cosmologically stable if it is made to be much lighter than $M_{W_R}$.
This brings in issues with vacuum stability, similar to the case of the SM~\cite{Weinberg:1976pe,Linde:1975sw},
and may require fine-tuning of couplings in the LRSM~\cite{Basecq:1988cv,Maiezza:2016bzp,Nemevsek:2016enw},
therefore we do not pursue this option any further.
}.
Without fine-tuning the flavor structure of $V_{\rm PMNS}^R$, the Feynman diagram in Fig.~\ref{fig:FeynDiagN1}
shows the dominant interaction for $N_1$ to couple with the charged leptons and quarks. 

If heavy enough, $N_1$ could decay into an electron and a charged pion, and it would not be cosmologically 
stable, unless the $W_R$ boson is ultra heavy, near the GUT scale~\footnote{One may consider the PMNS coupling 
of $N_1$ to $\tau$ only, which would prevent the tree-level decay to $e \pi$ and allow for slightly heavier $N_1$.}.
In this work, we are interested in finding lower (and upper) bounds of $W_R$, and we will focus on 
the $N_1$ dark matter mass below $\sim 100\,$MeV, where there is a wide portion of parameter space 
for $N_1$ to be sufficiently stable.
To be more specific, the upper bound on $N_1$ mass can be written as
\begin{equation}\label{eq:N1massupperbound}
  \begin{split}
  m_{N_1} &< \max \left( \left(\frac{96 \pi^3}{\tau G_F^2} \right)^{1/5}
  \left( \frac{M_{W_R}}{M_{W}} \right)^{4/5}, m_\pi \right)
  \simeq m_\pi \times \max \left( \left( \frac{M_{W_R}}{10^{10} \text{GeV}} \right)^{4/5}, 1 \right) \, ,      
  \end{split}
\end{equation}
where we require dark matter decay rate $\tau^{-1} \lesssim 10^{-50} \text{ GeV}$, which corresponds to a typical bound from 
cosmic ray positrons and X-ray searches.
As for laboratory constraints, the mass of the $W_R$ boson is chiefly bounded by searches at the 
Large Hadron Collider (LHC).
For RH neutrinos below a few 10 of GeV, the signal from $W_R \to \ell N_1$ looks like a very
energetic charged lepton and missing energy, because $N_1$ escapes detection.
Recasting the generic $W' \to \ell \slashed E$ searches~\cite{Nemevsek:2018bbt}, the current bounds 
set the LR scale to be $M_{W_R} \gtrsim 5\,$TeV, depending on the flavor of the charged lepton.
This channel is particularly clean and may probe the scales up to 37 TeV at a future 100 TeV 
collider~\cite{Nemevsek:2023hwx}.

With such super-weak interactions, mediated by the Feynman diagram in Fig.~\ref{fig:FeynDiagN1}, the 
RH neutrinos thermalize in the early universe and decouple at temperatures
\begin{equation} \label{eq:TDecN}
  T_d \sim 1\,{\rm MeV} \left( \frac{M_{W_R}}{M_W} \right)^{4/3} \gtrsim 300\,{\rm MeV} \, ,
\end{equation}
where in the second step we apply the existing lower bound on $M_{W_R}$ from the LHC.
Comparing with Eq.~\eqref{eq:N1massupperbound}, we find that
\begin{equation}
  T_d > m_{N_1} \, ,
\end{equation}
always holds.
We will assume that the reheating temperature of the early universe was sufficiently high, such 
that all the RH neturinos (and other particles in the LRSM) were once kept in thermal equilibrium by 
gauge interactions.
The above comparison implies that $N_1$ must decouple when it was still ultra relativistic, similar 
to the decoupling of SM neutrinos.
As will be sharpened in section~\ref{sec:nodilution}, this leads to a severe dark matter 
overproduction problem and requires a non-standard cosmology after the freeze out.

%
\subsubsection{Neutrino mass contributions}

Neutrino masses in the minimal LRSM come from two sources. 
First, the Dirac neutrino mass term together with the Majorana mass for $N$, generated by spontaneous 
symmetry breaking, can give mass to the active neutrinos through the type-I seesaw mechanism.
In addition, the model features another contribution through the type-II seesaw. 
It comes from the VEV of the left-handed scalar triplet $\langle \Delta_L^0 \rangle = v_L/\sqrt{2}$ 
and the $\lambda$ Yukawa coupling term in Eq.~\eqref{eq:Yukawa}.
The $v_L$ condensate originates from terms in the scalar potential of the form, e.g., 
${\rm Tr}(\Delta_L \Phi \Delta_R \Phi^\dagger)$, which is a tadpole term for $\Delta_L$ after $\Phi$ 
and $\Delta_R$ have obtained their condensates. 

The full neutrino mass matrix in the model is then given by
\begin{align} \label{eq:Seesaw}
  M_\nu &= - M_D^T M_N^{-1} M_D + M_L \, ,
  &
  M_D &= \frac{v}{\sqrt 2} \left( \cos \beta \, Y_l  + \sin \beta e^{-i\alpha} \, \tilde Y_l \right) \, ,
  \\
  M_N &= \frac{v_R}{\sqrt 2} \, Y_{\Delta_R} \, ,
  &
  M_L &= \frac{v_L}{\sqrt 2} \, Y_{\Delta_L} \, .
\end{align}
Accommodating the neutrino masses and mixings, needed to explain neutrino masses and fit the neutrino 
oscillations, is one of the primary motivations for considering the LRSM as a plausible BSM theory.
We will use it as an important guiding principle when exploring cosmological aspects of the model. 

\subsubsection{
\texorpdfstring{$W-W_R$ gauge boson mixing}{WL-WR gauge boson mixing}}\label{sec:WWRmix}

Because the scalar bidoublet $\Phi$ transforms under both $SU(2)_L$ and $SU(2)_R$, its VEVs allow 
for a mixing between the $W$ and $W_R$ gauge bosons. 
The corresponding mass terms take on the form
\begin{equation}
  \mathcal{L}_W = - 
  \begin{pmatrix}
    A_{L \mu}^- & A_{R \mu}^-
  \end{pmatrix} \frac{g^2}{2}
  \begin{pmatrix}
    \frac{1}{2} (v^2 + 2 v_L^2) & -  v^2 \sin\beta \cos\beta e^{-i\alpha} 
    \\ 
    - v^2 \sin\beta \cos\beta e^{i\alpha} & \frac{1}{2}(v^2+ 2 v_R^2)    
  \end{pmatrix}
  \begin{pmatrix}
    A_L^{\mu +} \\ A_R^{\mu +}     
  \end{pmatrix}
  \, ,
\end{equation}
where $A_{L,R}^\pm$ are the gauge bosons in the flavor basis.
After diagonalization, the mass eigenstates are the SM-like $W$ boson, which is a linear
superposition of mostly $A_L$ and a small admixture of $A_R$ that mediates RH currents.
Vice-versa, the $W_R$ mass eigenstate is mostly right-handed.
Phenomenologically, the relevant scales in the LRSM need to hierarchical, such that 
$v_R \gg v \gg v_L$ and the $W-W_R$ mixing is approximated by
\begin{equation}\label{eq:xiLR}
  \xi_\text{LR} \simeq \sin \beta \cos \beta e^{i \alpha} \, \left( \frac{v}{v_R} \right)^2
           \simeq \sin (2 \beta) e^{i \alpha} \, \left(\frac{M_{W}}{M_{W_R}} \right)^2 \ ,
\end{equation}
where in the last step we used $M_{W}^2 \simeq g^2 v^2/4$ and $M_{W_R}^2 \simeq g^2 v_R^2/2$.
As we will see, $\xi_\text{LR}$ is one of the key LRSM parameters for resolving the dark matter 
re-population issue.


Clearly $\xi_\text{LR}$ is suppressed due to the small mass ratio $(M_W/M_{W_R})^2$.
The magnitude of $\xi_\text{LR}$ also depends on $\tan \beta$, i.e. the ratio of the two VEVs from the bi-doublet
for which there exist an upper bound from the perturbativity of Yukawa couplings.
To make the point, we consider the quark mass generation in the minimal LRSM here and explain the
logic by approximating with the third family only, which has the largest Yukawa couplings.
For a single generation, the top and bottom quark masses are given by
\begin{align}
  m_t &= \frac{v}{\sqrt{2}} \left(Y_q \cos\beta + \tilde Y_q \sin\beta e^{-i\alpha} \right) \, , 
  &
  m_b &= \frac{v}{\sqrt{2}} \left(Y_q \sin\beta e^{i\alpha} + \tilde Y_q \cos\beta \right) \, .
\end{align}
These can be inverted and solved for the Yukawa couplings
\begin{align}
  Y_q \, e^{i\alpha} &= \frac{\sqrt{2}}{v \cos2\beta} \left( m_t \, e^{i\alpha} \cos\beta - m_b \, \sin\beta \right) \, , 
  &
  \tilde Y_q         &= \frac{\sqrt{2}}{v\cos2\beta} \left( - m_t \, e^{i\alpha} \sin\beta + m_b \, \cos\beta \right) \, .
\end{align}
By requiring $|Y_q|$ and $|\tilde Y_q|$ to take on perturbative values ($\lesssim 3$) for 
$\alpha\in[0, 2\pi)$, we get the following allowed range for $\tan \beta$
\begin{equation}\label{eq:tanbetarange}
  \tan\beta< -1.24 \quad \cup \quad -0.77< \tan \beta < 0.77 \quad \cup\quad \tan \beta > 1.24 \, .
\end{equation}
In other words, the large difference between the top and bottom quark masses forbids the angle $\beta$
to be close to $\pm\pi/4$, where $\cos2\beta$ approaches to zero blowing up $Y_q$ and $\tilde Y_q$.
See~\cite{Maiezza:2010ic} for more details and a full numerical study with 
three generations.

%
\subsubsection{The Majorana Higgs} \label{sec:Delta}

As mentioned earlier, the gauge symmetry breaking from LRSM to the SM is triggered by the VEV of 
the $SU(2)_R$ scalar triplet $\Delta_R$.
Its components $\delta_R^\pm$, and the imaginary part of $\delta_R^0$, become the longitudinal components 
of the $W_R^\pm$ and $Z'$ bosons, respectively. 
The real part of $\delta_R^0$ is the ``Higgs boson'' for this step of symmetry breaking, a massive 
propagating particle that reveals the nature of spontaneous breaking.
We denote the properly normalized physical state as $\Delta$, where 
\begin{equation}
\Delta \equiv \sqrt{2}\delta_R^0 - v_R \ .
\end{equation}
Because $v_R$ violates lepton number and serves as the source of Majorana neutrino mass, $\Delta$ is referred to as the 
Majorana Higgs~\cite{Nemevsek:2016enw}.
The doubly charged component $\Delta_R^{++}$ is left over as another physical state.

As will be discussed section~\ref{sec:HiggsDilution}, $\Delta$ can also play a crucial role of dilution for 
addressing the dark matter relic density. 
Here, we list its interactions that are relevant for understanding its role in the early universe.
The $\Delta$ is the excitation above the VEV $v_R$, which is mostly responsible for the mass generation for 
the right-handed neutrinos $N$ and $W_R^\pm$, $Z'$ gauge bosons. 
The corresponding couplings can be derived by shifting $v_R\to v_R (1+ \Delta/v_R)$,
\begin{equation}
  \mathcal{L} = -\frac{m_{N}}{v_R} \bar N N \Delta 
                + 2 \frac{M_{W_R}^2}{v_R} W_{R\mu}^+ W_R^{-\mu} \Delta 
                + \frac{M_{Z'}^2}{v_R} Z'_\mu Z'^\mu \Delta \, ,
\end{equation}
where we keep the interaction terms linear in $\Delta$, which are useful for calculating its decay rates.
The $\Delta-N$ coupling is diagonal in the mass basis of $N$.

Another important parameter that controls the decay rates of $\Delta$ is its mixing with the Higgs boson, 
$\theta_{\Delta h}$.
In the presence of such mixing, $\Delta$ can decay into all the SM particles that the Higgs boson couples to. 
In particular, when the mass of $\Delta$ is much above the electroweak scale, it mainly decays into $W^+W^-$, 
$Z Z$ and $h h$,  with a ratio of $2 : 1 : 1$, as dictated by the equivalence principle.
The scalar potential terms that couple $\Delta$ to Higgs are 
$\alpha_1 {\rm Tr}(\Phi^\dagger \Phi){\rm Tr}(\Delta_R^\dagger \Delta_R)+[\alpha_2 {\rm Tr}(\Phi^\dagger \tilde \Phi){\rm Tr}(\Delta_R^\dagger \Delta_R) + {\rm h.c.} ] + 
\alpha_3 {\rm Tr}(\Phi^\dagger \Phi\Delta_R^\dagger \Delta_R)$~\cite{Deshpande:1990ip}. 
After the right-handed triplet $\Delta_R$ develops the VEV, but before the electroweak symmetry breaking, 
these terms allow $\Delta$ to decay into a pair of SM Higgs bosons, as well as the would-be Goldstone bosons 
that eventually become the longitudinal components of the $W$ and $Z$ bosons.

%
%
\section{General Dilution mechanism vs. large scale structure} \label{sec:DMBasics}

We review the dark matter dilution mechanism under the sudden decay approximation~\cite{Scherrer:1984fd}, 
which is a useful tool for exploring the late decay of long-lived particles in the early universe.
This approximation allows us to analytically derive the important parametrical dependence in relevant quantities, 
such as the final dark matter relic density $\Omega_X$, and the reheating temperature $T_{\rm RH}$ immediately 
after the decay of the dilutor.
To keep the discussion here as general as possible, we call here the dark matter particle $X$ and the 
cosmologically ``long-lived'' particle for entropy dilution $Y$. 
We will assign their identities within the LRSM ($X\to N_1$, $Y \to N_2, \Delta$) in the next section, when we discuss the 
concrete dark matter dilution scenarios.

\subsection{Relativistic freeze out and overproduction problem} \label{sec:nodilution}

Consider the dark matter $X$, which freezes out from the SM thermal plasma relativistically. 
The yield $Y_X = n_X/s$ is then defined as the ratio of number density $n_X$ to the total entropy density $s$
of the SM plasma, and is given by
\begin{equation}\label{app1}
  Y_X = \frac{135 \, \zeta(3)}{4\pi^4 g_*(T_{\rm fo})} \, ,
\end{equation}
where we assumed that $X$ is a Majorana fermion with 2 degrees of freedom.
The $T_{\rm fo}$ is the photon temperature when $X$ freezes out, and $g_*(T_{\rm fo})$ counts the corresponding 
number of relativistic degrees of freedom in the universe in the plasma. 
Because most of our discussion will be restricted to temperatures above the MeV scale for successful big-bang
nucleosynthesis (BBN), we will not distinguish $g_*(T)$ and $g_{*S}(T)$ hereafter. 
If nothing else happened after the freeze out, $Y_X$ would be a conserved quantity, and the dark matter relic 
density today would be
\begin{equation}
  \Omega^0_X = \frac{m_X \, Y_X \, s_0}{\rho_0} \simeq 2.6 \, \left( \frac{m_X}{1\,\rm keV} \right) 
  \left( \frac{100}{g_*(T_{\rm fo})} \right) \ ,
\end{equation}
where $s_0 = 2891.2 \text{ cm}^{-3}$ is the entropy density in the universe today, and 
$\rho_0 = 1.05 \times 10^{-5} h^{2}\,{\rm GeV/cm^3}$ represents today's critical density with $h = 0.67$~\cite{Zyla:2020zbs}.
In contrast, the {\it Planck} experiment observes that the value of $\Omega_\text{dark matter}$ is 0.26~\cite{Aghanim:2018eyx}. 
Because $m_X$ is constrained to be heavier than several keV due to various warm dark matter constraints,
the above result creates the dark matter overproduction problem.

\subsection{Entropy dilution mechanism with a long-lived particle}

To address the issue of overproduction, we introduce a dilutor particle $Y$. 
For simplicity, we assume it is also a Majorana fermion that freezes out relativistically and has a similar 
yield as the dark matter before decaying away, mostly into the SM particles. 
To achieve sufficient dilution, $Y$ must dominate the total energy density of the universe (as matter) before 
it decays away and dumps most of its energy (or entropy) into the SM sector. 

In the sudden decay approximation, we have
\begin{equation}\label{app3}
  \tau_Y^{-1} = H_{\rm before} = H_{\rm after} \, ,
\end{equation}
where $\tau_Y$ is the lifetime of $Y$, and $H_{\rm before,\, after}$ are the Hubble parameters 
($H\equiv \sqrt{8\pi G_N\rho/3}$) immediately before and after the decay, respectively. 

Before the decay of $Y$, the universe is dominated by the energy density of non-relativistic massive $Y$ particles,
\begin{equation}
  \rho = \rho_Y =  Y_Y s_{\rm before} m_Y \ ,
\end{equation}
where $s_{\rm before}$ is the total entropy density of relativistic species before the decay.
Here we assume that $Y$ experiences a similar relativistic freeze out as dark matter and $Y_Y$ is the same as 
$Y_X$ given in Eq.~\eqref{app1}.
Note that the universe is already matter dominated right before $Y$ decays. 
However, this does not prevent us from defining $Y$ as the ratio of $n_Y$ to the relativistic entropy density 
$s$, and $Y_Y = n_Y/s$ remains conserved in the time window between the freeze-out and $Y$'s decay.
With these inputs, the first of Eq.~\eqref{app3} leads to
\begin{equation}\label{app5}
  s_{\rm before} = \frac{\pi^3 g_*(T_{\rm fo})}{90 \zeta(3)} \frac{M_{\rm pl}^2}{m_Y \tau_Y^2} \, .
\end{equation}

Assuming that the decay of $Y$ takes no time, the energy density of $Y$ immediately before its decay is equal 
to the radiation energy density immediately after.
The latter is related to the corresponding ``reheating'' temperature $T_{\rm RH}$ of the SM plasma,
\begin{equation}
  \rho = \rho_R = \frac{\pi^2}{30} g_*(T_{\rm RH}) T_{\rm RH}^4 \, .
\end{equation}
The second equation of Eq.~\eqref{app3}, $\tau_Y^{-1} = H_{\rm after}$, leads to
\begin{equation}\label{app4}
  T_{\rm RH}    \simeq 0.6 \, g_*(T_{\rm RH})^{-1/4} \sqrt{\frac{M_{\rm pl}}{\tau_Y}} 
                \simeq \frac{0.93 \, {\rm MeV}}{g_*(T_{\rm RH})^{1/4}} \sqrt{\frac{1\,\rm sec}{\tau_Y}} \, ,
\end{equation}
where $M_{\rm pl} = \sqrt{1/G_N} =1.2\times10^{19}\,$GeV is the Planck constant.
The entropy density of the SM plasma immediately after $Y$ decay can then be calculated in terms of $T_{\rm RH}$,
\begin{equation}\label{app6}
  s_{\rm after} = \frac{2\pi^2}{45} g_*(T_{\rm RH}) T_{\rm RH}^3 \ .
\end{equation}
With Eqs.~\eqref{app5} and \eqref{app6} we can derive the dilution factor $\mathcal{S}$,
\begin{equation}\label{eq:dilutionfactorS}
  \mathcal{S} \equiv \frac{s_{\rm after}}{s_{\rm before}} 
  \simeq \frac{0.7 \, g_*(T_{\rm RH})^{1/4}}{g_*(T_{\rm fo})} \frac{m_Y \sqrt{\tau_Y}}{\sqrt{M_{\rm pl}}} \ .
\end{equation}

The diluted relic density of $X$ today is given by
\begin{equation}\label{eq:RelicDensityAfterDilution}
\begin{split}
  \Omega_X = \frac{\Omega_X^0}{\mathcal{S}} &\simeq \frac{0.72 g_*(T_{\rm RH})^{1/4}}{g_*(T_{\rm fo})} 
  \frac{m_Y \sqrt{\tau_Y}}{\sqrt{M_{\rm pl}}} 
  \\
  &\simeq 0.26 \left( \frac{m_X}{1\,\rm keV} \right) \left(\frac{2.2\,\rm GeV}{m_Y}\right) 
  \sqrt{\frac{1\,\rm sec}{\tau_Y}} \ .
\end{split}
\end{equation}
This is the standard dark matter dilution mechanism that has been employed in various contexts for addressing the 
dark matter relic density.

%
\subsection{Dilutor to dark matter decay}

We recently pointed out~\cite{Nemevsek:2022anh} new opportunities to test the dark matter dilution mechanism. 
We showed that the re-population of dark matter in $Y$ decays leaves an imprint on structure formation, and gets 
constrained by the existing data (or gives a characteristic signal upcoming data).
While in~\cite{Nemevsek:2022anh} we worked in a largely model independent way, in a concrete UV realization of 
such mechanisms, the model predicts not only the relic density, but also the spectrum -- the phase space 
distribution of dark matter with a primary and secondary component.
To account for all these possibilities, we consider the following decay channels of $Y$, where it can decay into 
SM particles as well as dark matter $X$,
\begin{align}\label{eq:decaychannels}
  Y &\to SM \, , & Y &\to n X \, (\ +\ m \ SM \ ) \ ,
\end{align}
where $n, m \in \mathbb Z^+$ count the multiplicity of $X$ and SM particles, respectively, in the final states. 
The first decay channel is desired for dumping entropy into the visible sector and dilutes the primordial thermal 
population of $X$.
If this were the only final state of $Y$ decay, the resulting $X$ would remain a purely thermal distribution with
a temperature $T_X$, which would be relatively lower than the counterpart in the absence of dilution.

The other decay mode, whose branching ratio is assumed to be ${\rm Br}_X$, produces a secondary non-thermal population 
of $X$ that also contributes to the final dark matter relic abundance. 
The bracket in~\eqref{eq:decaychannels} also includes the possibility that this second decay channel is completely dark,
without any ``SM'' in the final state. 
In the absence of extended dark sectors, the branching ratio of the first channel is simply $1 - {\rm Br}_X$.
With a nonzero ${\rm Br}_X$, the final dark matter relic density becomes
\begin{equation}\label{eq:RelicDilute} 
  \Omega_X \simeq 0.26 \, \left(1 + n {\rm Br}_X \right) \left( \frac{m_X}{1\,\rm keV} \right) 
  \left( \frac{2.2\,\rm GeV}{m_Y} \right) \sqrt{\frac{1\,\rm sec}{\tau_Y}} \, .
\end{equation}
Requiring $\Omega_X$ to agree with the {\it Planck} measured value fixes $\tau_Y$ in terms of $m_X$ and $m_Y$, 
we can rewrite Eq.~\eqref{app4} as
\begin{equation}\label{app9}
  T_{\rm RH} \simeq \frac{0.4\,{\rm MeV}}{g_*(T_{\rm RH})^{1/4}} \frac{m_Y}{10^6 m_X} \ .
\end{equation}
Because $g_*$ is always larger than 1, and for successful big-bang nucleosynthesis to work, for which 
$T_{\rm RH} \gtrsim 1 \text{ MeV}$ is needed, the dilutor $Y$ must be heavier than dark matter $X$ by 
a factor of at least a million.

Before deriving and solving the equation for dark matter phase space distribution, we first give a qualitative 
discussion and introduce an important temperature relevant for large scale structure of the universe.
Immediately after the $Y\to n X + m \, SM$ decay, each secondary $X$ particle roughly carries the energy of 
$m_Y/(m + n)$. 
Under the sudden decay approximation, the corresponding temperature of the universe is given by $T_{\rm RH}$ 
in Eq.~\eqref{app4}.
The velocity of $X$ particles will then redshift with the expansion of the universe.
After a while the $X$ particles start to turn non-relativistic when the energy drops to around their mass.
This requires the scale factor of the universe to grow by a factor of
\begin{equation}
  \frac{a_{\rm NR}}{a_{\rm RH}} \simeq \frac{m_Y}{(m+n) \, m_X} \, .
\end{equation}
The corresponding temperature $T_{\rm NR}$ can be found with entropy conservation in the SM sector 
\begin{equation}
  g_{*S}(T_{\rm NR}) T_{\rm NR}^3 a_{\rm NR}^3 = g_*(T_{\rm RH}) T_{\rm RH}^3 a_{\rm RH}^3 \, ,
\end{equation}
which leads to
\begin{equation} \label{eq:TNR}
  T_{\rm NR} = T_{\rm RH} \left(\frac{g_*(T_{\rm RH})}{g_*(T_{\rm NR})} \right)^{1/3} 
  \frac{a_{\rm RH}}{a_{\rm NR}} \simeq 0.25 \, {\rm eV} \, n \, g_*(T_{\rm RH})^{\frac{1}{12}} \, .
\end{equation}
In the second step we used Eq.~\eqref{app9} and the late-time value for $g_{*S}(T_{\rm NR}) = 3.91$, valid 
for $T_{\rm NR}$ well below the electron mass.

Note that $T_{\rm NR}$ defines the time when both the primordial $X$ particles and the secondary ones from $Y$ 
decay have become matter-like.
Dialing the clock back to temperatures above $T_{\rm NR}$, the dark matter fluid is made out of the non-relativistic 
primordial and the relativistic secondary component. 
The energy density of the latter is more important at temperatures above $T_{\rm NR}/{\rm Br}_X$.
In this regime, the overall $X$ fluid is relativistic and features a large pressure, which can interrupt the 
regular logarithmic growth of matter density perturbations in $X$. 
This suppresses the matter power spectrum $P(k)$ for wavelengths of the perturbation smaller than the Hubble
radius at temperature equal to $T_{\rm NR}/{\rm Br}_X$. 
The resulting $P(k)$ may then potentially disagree with the LSS measurements, unless ${\rm Br}_X \ll 1$.

%
\subsection{Phase space distribution of dark matter} \label{sec:psd}

Let us go beyond the sudden decay approximation and derive the equations governing the dark matter phase 
space distribution.
Because the temperature $T_{\rm NR}$, when the secondary dark matter particles from dilutor decay turn 
non-relativistic, is found to be rather low, they act as a hot dark matter component for a period of 
time that overlaps with the observational data.
Consequently, they may suppress the large and small scale structures, which in turn allows one to
derive a powerful constraint on the dilutor $\to$ dark matter branching ratio using cosmological data from 
the Sloan Digital Sky Survey (SDSS)~\cite{Nemevsek:2022anh}.

Without loss of generality, we write the dilutor to dark matter decay channel as
\begin{equation} \label{eq:Ydecay}
  Y \to X + 2 + 3 + \cdots + N = n \, X + m \, \text{SM} \, ,
\end{equation}
where particles $2, 3, \dots, N = m + n$ represent either the SM or additional $X$ particles in the 
final state.
As before, we assume that $n X$ particles are produced in this decay.
The phase space distribution for the produced $X$ is governed by the Liouville's equation
that relates the phase space distribution functions of dark matter $f_X$ and the dilutor $f_Y$
\begin{align} \label{eq:Liouville}
  &\left( \frac{\partial}{\partial t} - H \frac{|\vec p_X|^2}{E_X} \frac{\partial}{\partial E_X} \right) f_X \left(E_X, t \right) = 
  \int \frac{\text{d}^3\vec{p}_Y}{(2\pi)^3} \frac{1}{2E_Y} f_Y(E_Y, t) \, A \, ,
  \\
  & A = \frac{1}{2E_X}\prod_{i = 1}^N \int \frac{\text{d}^3\vec{p}_i}{(2\pi)^3} 
  \frac{1}{2E_i} (2\pi)^4 \delta^4\left(p_Y - p_X - \sum_i p_i\right) \left| \mathcal{M} \right |^2 \, .
\end{align}
Hereafter we work in the ultra-relativistic $X$ limit (from $Y$ decay) and approximate $|\vec{p}_X| \simeq E_X$.
$\mathcal{M}$ is the decay matrix element related to the decay in~\eqref{eq:Ydecay} and $A$ 
can be rewritten in terms of the partial decay rate of $Y \to X + \cdots + N$ in its rest frame as
\begin{equation}
  \Gamma_{Y \to X} = \frac{1}{2m_Y} \int \frac{\text{d}^3\vec{p}_X}{(2\pi)^3} \, A = 
  \frac{1}{4 \pi^2 m_Y} \int \text{d} E_X \, E_X^2  A \ ,
\end{equation}
where it is assumed that the $X$ particles from $Y$ decay are ultra-relativistic throughout most of 
the phase space. 
Here it is useful to introduce a dimensionless spectral function $g(\omega)$, which satisfies
\begin{align}
  n \, \frac{\text{d} \Gamma_{Y\to X}}{\text{d} \omega} &= \Gamma_{Y \to X} \, g\left( \omega \right) \, ,
  & 
  \int \text{d}\omega \, g(\omega) &= n \, ,
  &
  \omega &= \frac{E_X}{m_Y} \, .
\end{align}
It allows us to establish a relation between $A$ and $g$ and simplify the Liouville's equation to
\begin{equation} \label{eq:Liouville2}
\begin{split}
  \left( \frac{\partial}{\partial t} - H E_X \frac{\partial}{\partial E_X} \right) f_X(E_X, t) &= 
  \frac{4 \pi^2}{E_X^2} \Gamma_{Y\to X} g\left( \frac{E_X}{m_Y} \right) \int 
  \frac{\text{d}^3\vec{p}_Y}{(2\pi)^3} \frac{1}{2 E_Y} f_Y \left(E_Y, t \right) \, .
\end{split}
\end{equation}

Next, we assume that $Y$ had already turned non-relativistic when the decay occurs, i.e., $E_Y \simeq m_Y$.
This is a necessary condition for the dilution mechanism to work, because $Y$ is assumed to dominate the
energy content in the universe as a matter component.
It allows us to complete the $p_Y$ integral in~\eqref{eq:Liouville2} and express it with the number density, $n_Y(t) = \rho_Y(t)/m_Y$. $\rho_Y$ is the energy density of non-relativistic $Y$ particles. 

For the left-hand side of Eq.~\eqref{eq:Liouville2}, we change the variables of $f_X$ to $x \equiv E_X/T_X$
and $t$, where $T_X$ is the temperature of primordial $X$ warm dark matter defined above.
The appealing reason for such a change is that as long as the $X$ particles remain ultra-relativistic, 
the ratio $E_X/T_X$ stays invariant in the expanding universe. 
Using the identity,
\begin{equation}
  \left( \frac{\partial}{\partial t} - H E_X \frac{\partial}{\partial E_X} \right) f_X \left( E_X, t \right) = 
  \frac{\partial}{\partial t} f_X(x, t) \, ,
\end{equation}
we finally obtain the phase space equation for secondary dark matter from dilutor decay,
\begin{equation}
  \frac{T_X^3}{2\pi^2} x^2 \frac{\partial}{\partial t} f_X(x, t) = 
  n_Y(t) \Gamma_{Y \to X} \frac{T_X}{m_Y} g \left( \frac{T_X}{m_Y}x \right) \, .
\end{equation}

This equation is to be solved along with the following set of energy density Boltzmann equations
for $X$, $Y$ and the SM particles
\begin{align}
  \dot \rho_Y + 3 H \rho_Y &= - \Gamma_Y \rho_Y \, , 
  \\
  \dot \rho_X + 4 H \rho_X &= y {\rm Br}_X \Gamma_Y \rho_Y \, , 
  \\
  \dot \rho_{\rm SM} + \left(4 H - \frac{\dot g_*}{3g_*} \right) \rho_{\rm SM} &= 
  \left(1 - y {\rm Br}_X \right) \Gamma_Y \rho_Y \, , \label{eq:8}
\end{align}
where $\rho_{\rm SM}$ is the energy density carried by relativistic visible particles and 
$H^2 = 8\pi G_N (\rho_Y + \rho_X+\rho_{\rm SM})/3$ is the Hubble parameter. 
This set of equations applies for non-relativistic $Y$, while the $X$ population remains ultra-relativistic.
%

Using $T_X$ to keep track of time, the phase space function $f_X$ at late times can
be solved
\begin{equation} \label{eq:fX}
\begin{split}
  f_X(x) &= \frac{1}{e^x + 1} + \frac{2\pi^2}{x^2} \frac{\Gamma_Y {\rm Br}_X}{m_Y^2} 
  \int_{T_{\text{fin}}}^{T_\text{ini}} 
  \frac{\text{d} T_X}{T_X} \, \frac{\rho_Y}{T_X^2 H} \, g \left(\frac{T_X}{m_Y} x \right) \ .
\end{split}
\end{equation}
The first term of~\eqref{eq:fX} is the primordial Fermi-Dirac distribution of $X$ and the second is the 
non-thermal re-population of $X$.
The $T_X$ integral should cover the entire temperature range relevant for the production of the 
secondary component of dark matter.
It goes from an arbitrary high initial temperature $T_{\rm ini}$, which in practice we take 
$T_{\rm ini} = m_Y/10$ in order for $Y$ to already be non-relativistic, 
as assumed below~\eqref{eq:Liouville2}.
The final result is insensitive to the exact choice of $T_{\rm ini}$, because the 
$\rho_Y/(T_X^2 H)$ factor in the integrand is suppressed at higher $T_X$, as long as $T_X\ll m_Y$.
Moreover, the $g$ function cannot lift this suppression, because $T_X$ is bounded from above for
a given fixed $x$.

On the lower limit of integration, we need to go to sufficiently low temperatures, such that all of 
the $Y$ is depleted by decays into SM and $X$.
Because of the exponential suppression in $\rho_Y$, the exact $T_{\text{fin}}$ is also not relevant.
In practice we integrate down to temperatures corresponding to $t = 10 \, \tau_Y$, which sufficiently 
covers the entire period of non-relativistic $Y$ decay. 

\begin{figure}[t]
  \begin{center}
    \includegraphics[width=0.618\textwidth]{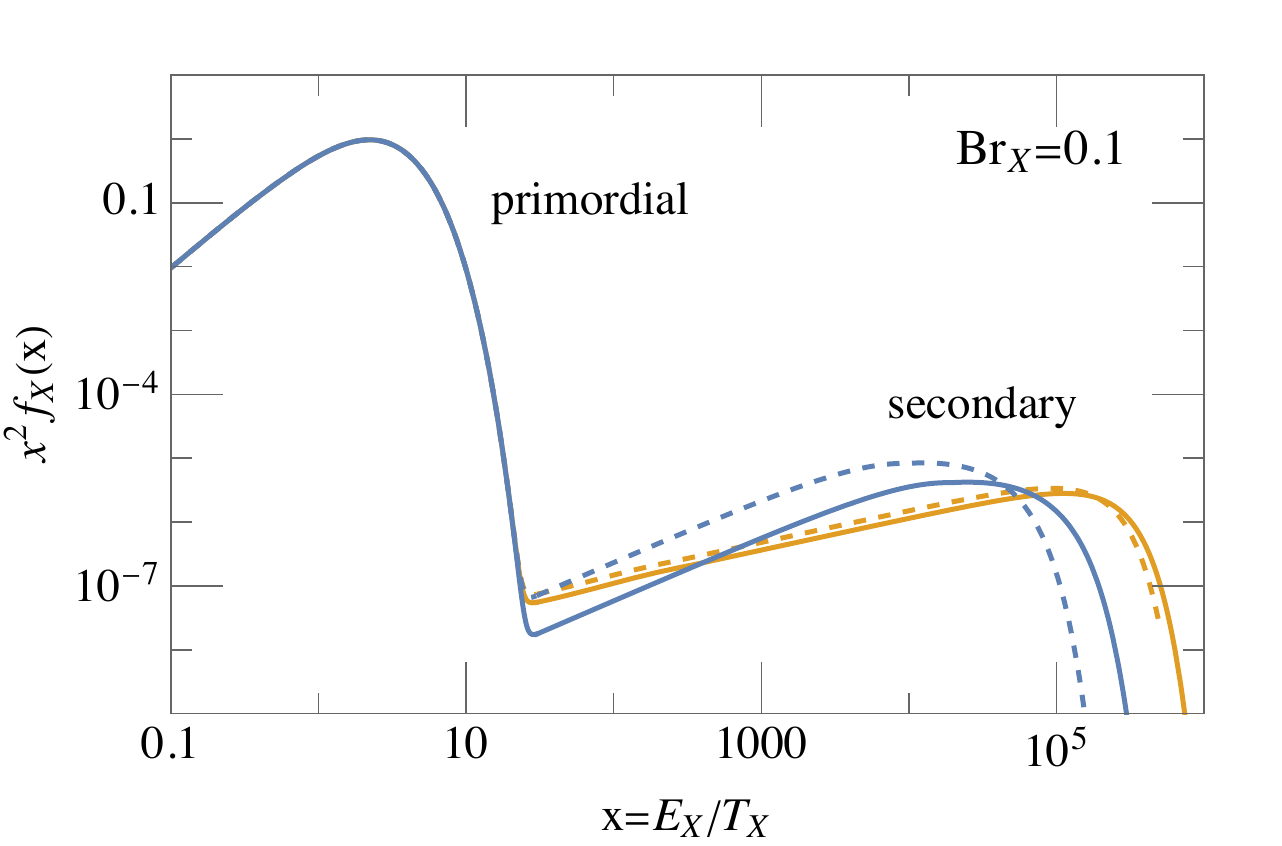}
  \end{center}
  \caption{
  Phase space distribution of ultra-relativistic dark matter species $X$. 
  The blue (orange) curve corresponds to the three-body (two-body) decay model listed in 
  TAB.~\ref{tab:models}.
  Solid (dashed) curve corresponds to $m_X = 10 \text{ keV}$, $m_Y = 100 \text{ GeV}$ 
  ($m_Y = 1 \text{ PeV}$), while $\text{Br}_X = 0.1$ for all the cases.
  }
\label{fig:fX}
\end{figure}

The resulting distributions $f_X$ are shown in FIG.~\ref{fig:fX}, which are plotted at late times after the 
dilution has completed.
The two options for two and three body decays correspond to two scenarios that are relevant for the minimal 
LRSM under consideration in this work.
\begin{enumerate}
  \item The case when $X$ is the lightest right-handed neutrino $N_1$ and $Y$ is a heavier $N_2$, which 
  undergoes a three-body decay into $N_1$ plus two charged leptons, mediated by the $W_R$ gauge boson;
  \item $Y$ is a long-lived scalar $\Delta_R$ with a partial decay width into two $N_1$.
\end{enumerate}
The corresponding $g$ functions and $n, y$ integrals are summarized in TAB.~\ref{tab:models}, where the 
masses of final-state charged leptons were neglected.

It is clear from FIG.~\ref{fig:fX} that the primary component of dark matter dominates at small $x$, while the 
secondary component features a significantly smaller occupancy, but carries more energy and thereby 
affects structure formation. 
On the plotted curves we kept fixed $m_X$ and $\text{Br}_X$ and took two different values of $m_Y$ to demonstrate
that the shape of the secondary component is roughly independent of the mass of the dilutor.
This follows from Eqs.~\eqref{eq:RelicDilute}, where the relic density requires the scaling $m_Y \sim 1/\sqrt{\tau_Y}$ and 
the reheating temperature is set by the Hubble time $T_{\rm RH} \sim \sqrt{H} \sim 1/\sqrt{\tau_Y}$. 
Immediately after $Y\to X$ decay, $E_X \lesssim m_Y$ and $T_X \sim T_{\rm RH}$. 
As a result, the kinematic endpoint $x_{\rm max}\sim m_Y/T_{\rm RH}$ is roughly held constant for fixed $m_X$, 
irrespective of the values of $m_Y$ or $\tau_Y$.

\begin{table}[ht]
  \centering
  \vspace{0.5cm}
  \begin{tabular}{|c|c|c|c|c|c|c|}
  \hline
   dark matter $X$ & dilutor $Y$ & $Y\to X$ decay            & $n$   & $g(\omega)$  & $y$ 
  \\
  \hline
  $N_1$ & $N_2$ & $N_2 \to N_1 + SM$    & 1     & $\,\,16 \omega^2 (3-4\omega) \, \theta \left(\frac{1}{2} - \omega \right) \,\,$ & $\frac{7}{20}$ 
  \\
  \hline
  $N_1$ & $\Delta$ & $\Delta \to N_1 N_1$   & 2     & $2 \delta (\omega - 1/2)$ & 1 
  \\
  \hline
  \end{tabular}
  \caption{
  The energy fraction distribution $g(\omega)$, taken away by dark matter $X$ in the rest frame of the decaying dilutor $Y$, and its integrals 
  $n = \int g, y = \int \omega g$. 
  In the context of LRSM, the first row corresponds to $N_2$ as dilutor which can undergo a three-body decay into a $N_1$ plus two charged leptons ($n=1$) and $\theta$ is the Heaviside unit step function.
  The second corresponds to the Majorana scalar boson $\Delta$ dilution scenario,
  where each $\Delta \to N_1N_1$ produces two dark matter states ($n=2$)
  and $\delta$ is the Dirac delta function.\vspace{0.2cm}}
 \label{tab:models}
\end{table}

In the case where the dilutor is a long-lived scalar $\Delta$ (see sec.~\ref{sec:Delta}), the kinematics is so simple that we can further
derive an explicit closed form for $f_X$.
Since it is a two body decay, $g$ is a Dirac-$\delta$ function and we can complete the $T_X$ integral 
in Eq.~\eqref{eq:fX} to obtain
\begin{align} \label{eq:Phi2XXPSD}
  x^2 f_X(x) &= \frac{x^2}{e^x + 1} + \frac{8\pi^2 {\rm Br}_X \Gamma_Y}{m_Y^2} 
  \left( \frac{\rho_Y}{T_X^2 H} \right)_{T_{X \star}} \, ,
  \\ \label{eq:Phi2XXPSDv2}
  &\simeq \frac{x^2}{e^x + 1} + 0.16 \text{ sec } {\rm Br}_X \left( \frac{10 \text{ keV}}{m_X} \right)^2
  \left( \frac{\rho_Y}{T_X^2 H} \right)_{T_{X_ \star}} \, ,
\end{align}
where we used the relic equation~\eqref{eq:RelicDilute}, set $\Omega_X = 0.26$ and approximated with 
small ${\rm Br}_X$ in the second step.
With such a simple expression in~\eqref{eq:Phi2XXPSDv2} we can understand the behaviour of 
$f_X$ in FIG.~\ref{fig:fX} that emerges from solving Eq.~\eqref{eq:fX}.
First of all, $g$ is a $\delta$ function in temperature, which essentially selects a particular moment in 
$T_{X \star} = m_Y/(2 x)$ for a fixed $x$ and thus completely removes any dependence on the 
boundary conditions $T_{\text{ini},\, \text{fin}}$.

Furthermore, we can derive the explicit dependence on $x$ for various moments in the expansion of the 
universe during the dilutors' decay.
For this, we only have to examine the $x$ dependence of the factor $\left(\rho_Y/T_X^2 H\right)_{T_{X_\star}}$ in Eq.~\eqref{eq:Phi2XXPSDv2}; the re-scaling with $\Omega_X$ and $m_X$ is trivial.
Note that for the purpose of this discussion, $Y$ is always non-relativistic and 
$\rho_Y \propto m_Y T_X^3$.
In the early stages of radiation domination we have $H \propto T_X^4$ and therefore
$\left(\rho_Y/T_X^2 H\right)_{T_{X_\star}} \propto m_Y/T_{X \star} \sim x$.
Once $Y$ starts to dominate, the Hubble parameter goes as $H \sim \sqrt{\rho_Y}$ and 
the relevant term goes as $\left(\rho_Y/T_X^2 H \right)_{T_{X_\star}} \propto \sqrt{m_Y/T_{X \star}} \sim \sqrt{x}$.
In both cases, $m_Y$ cancels out, that is why the orange curve in FIG.~\ref{fig:fX} remains 
almost identical for difference choices of $m_Y$.
Finally, large $x$ corresponds to low $T_{X\star}$ when all the $Y$ has decayed away and the
secondary part of $f_X$ is exponentially suppressed.
These $x$ dependencies explain the shape of the 2-body orange curves in the $x > 10$ region.

For the three-body decay, the spectral function $g$ is not as sharply peaked as the 
Dirac $\delta$, but has a maximum at $\omega = 1/2$ and similar considerations go through,
with transitions between different $x$ dependencies becoming less sharp.
The bottom-line is that, once we fix $Br_X$ and $m_X$, the $f_X$ does not dependent on the mass 
of the dilutor $Y$, the behavior of $f_X$ is roughly the same for different decay topologies
and it extends to large $x$, beyond the usual primary component.
In the following section we will examine how this behaviour translates onto the physical
matter power spectrum $P(k)$.

Let us emphasize that we assume the $X$ particles remain collisionless after the $Y$ decay throughout
this work.
This means that the imprint of the dark matter model (fundamental physics) on the phase space distribution 
is preserved until later times and can directly affect cosmological observations. 
We do not consider the possibility of having strong DM self-interactions. 
They could re-thermalize the dark sector and soften the above phase space distribution. 
At the same time, they would facilitate excessive dark matter production, which is adverse to the dilution 
mechanism considered here. 

%
\subsection{Imprint on the matter power spectrum} \label{sec:LSS}

\begin{figure}[ht]
  \begin{center}
    \includegraphics[width=0.75\textwidth]{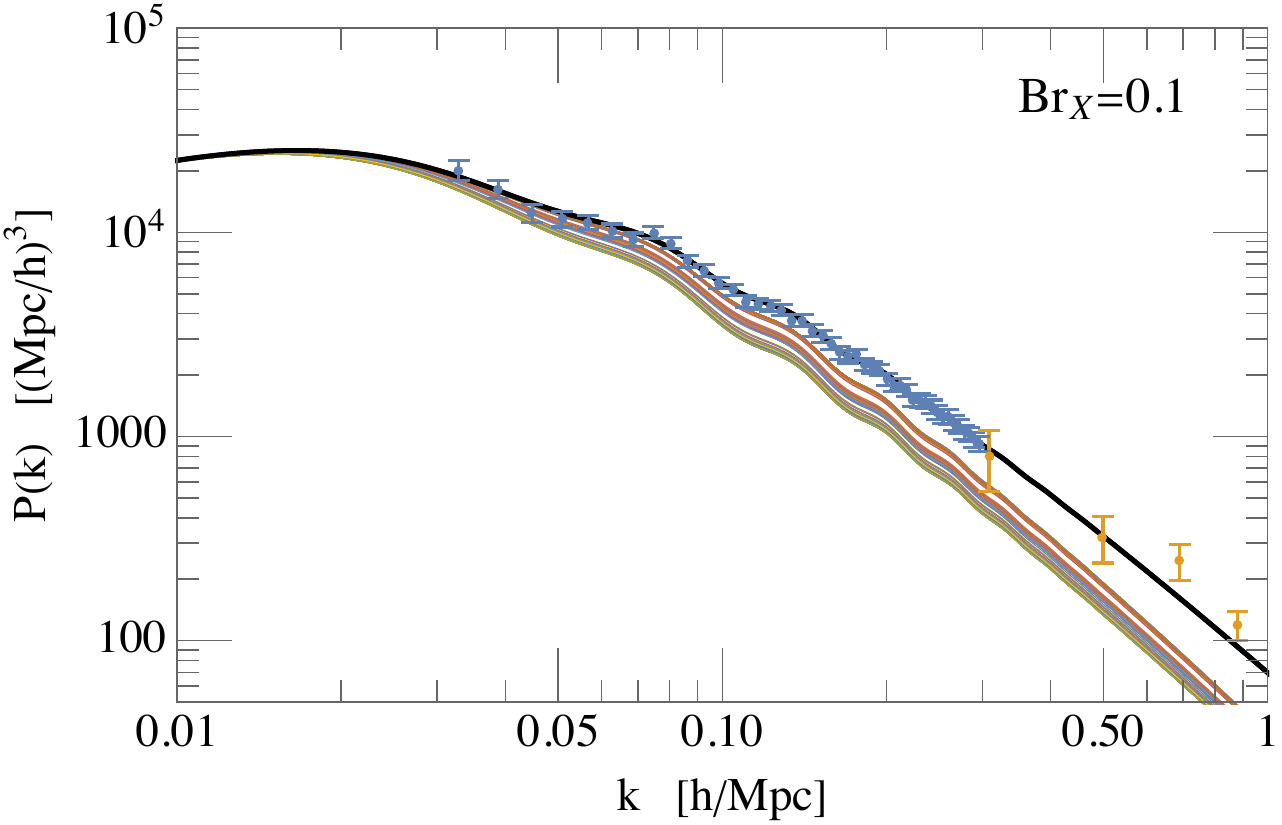}
  \end{center}
  \caption{Primordial matter power spectrum in standard $\Lambda$CDM (black solid curve) and a set of diluted dark matter models 
  listed in TAB.~\ref{tab:models} (colorful curves). 
  Like in FIG.~\ref{fig:fX}, we set ${\rm Br}_X = 0.1$. 
  Data points from SDSS DR7 LRG and Lyman-$\alpha$ observations are shown in blue and orange, respectively.}
\label{fig:pk}
\end{figure}

Let us turn to a quantitative numerical analysis in the parameter space of $m_X$ versus $m_Y$. 
For each point we first set the dilutor lifetime $\tau_Y$ using Eq.~\eqref{app9}. 
Next, we determine the phase space distribution $f_X$ with Eq.~\eqref{eq:fX} and evolve the density perturbations using the linear Boltzmann solver code 
{\tt CLASS}~\cite{Lesgourgues:2011re, Blas:2011rf, Lesgourgues:2011rh} to obtain the corresponding matter power spectrum $P(k)$. 
We scan over $200$ points in the mass range $m_X \in (1\,{\rm keV}, 1\,{\rm MeV})$ and 
$m_Y \in (1\,{\rm GeV}, 10^{16}\,{\rm GeV})$ for both decay channels considered in TAB.~\ref{tab:models}. 
The results are shown by the colored curves in FIG.~\ref{fig:pk}, where we set ${\rm Br}_X = 0.1$. The black 
solid curve is the fiducial $\Lambda$CDM model.

\begin{figure}[t]
  \begin{center}
    \includegraphics[width=0.48\textwidth]{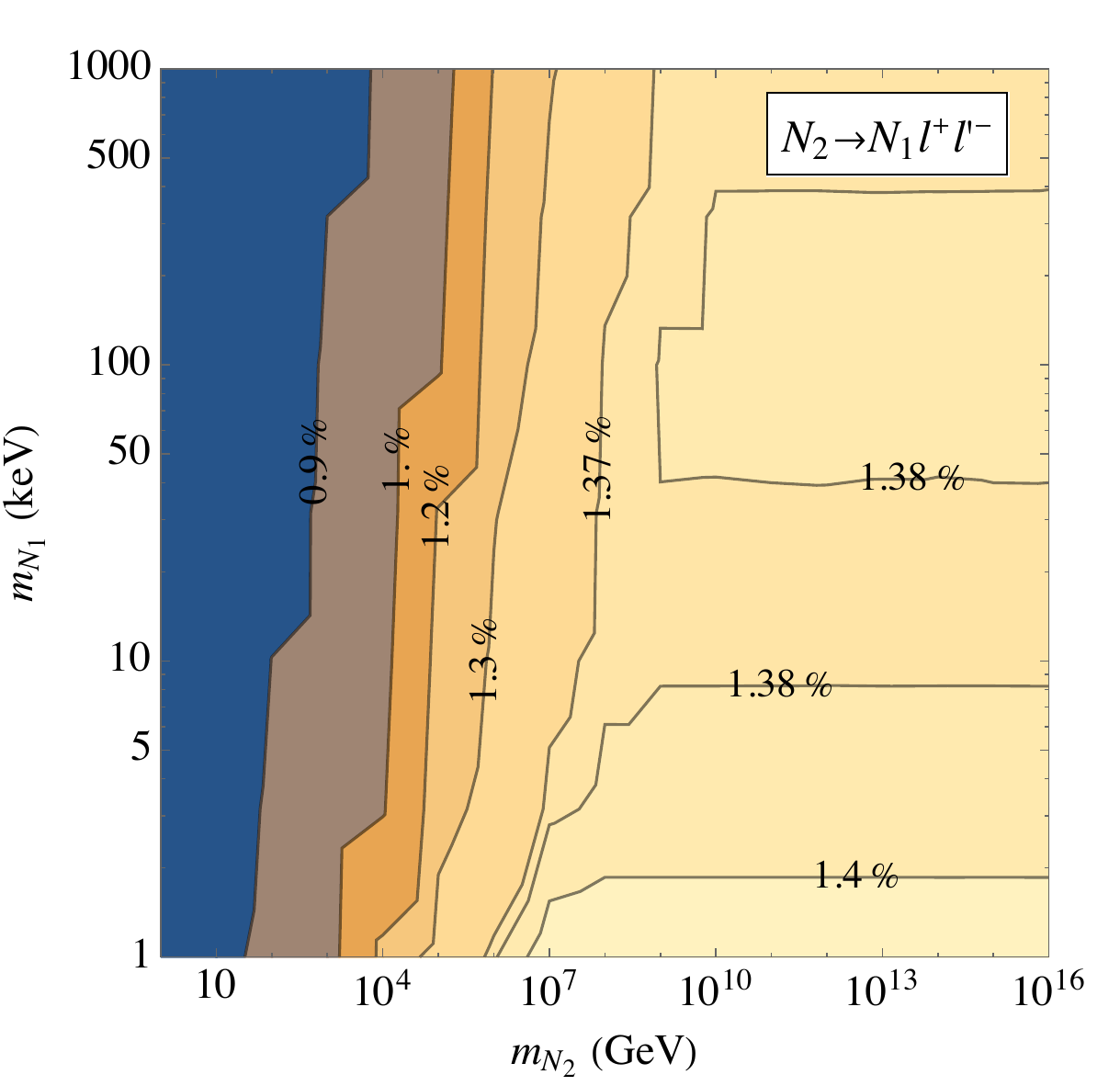} \quad
    \includegraphics[width=0.48\textwidth]{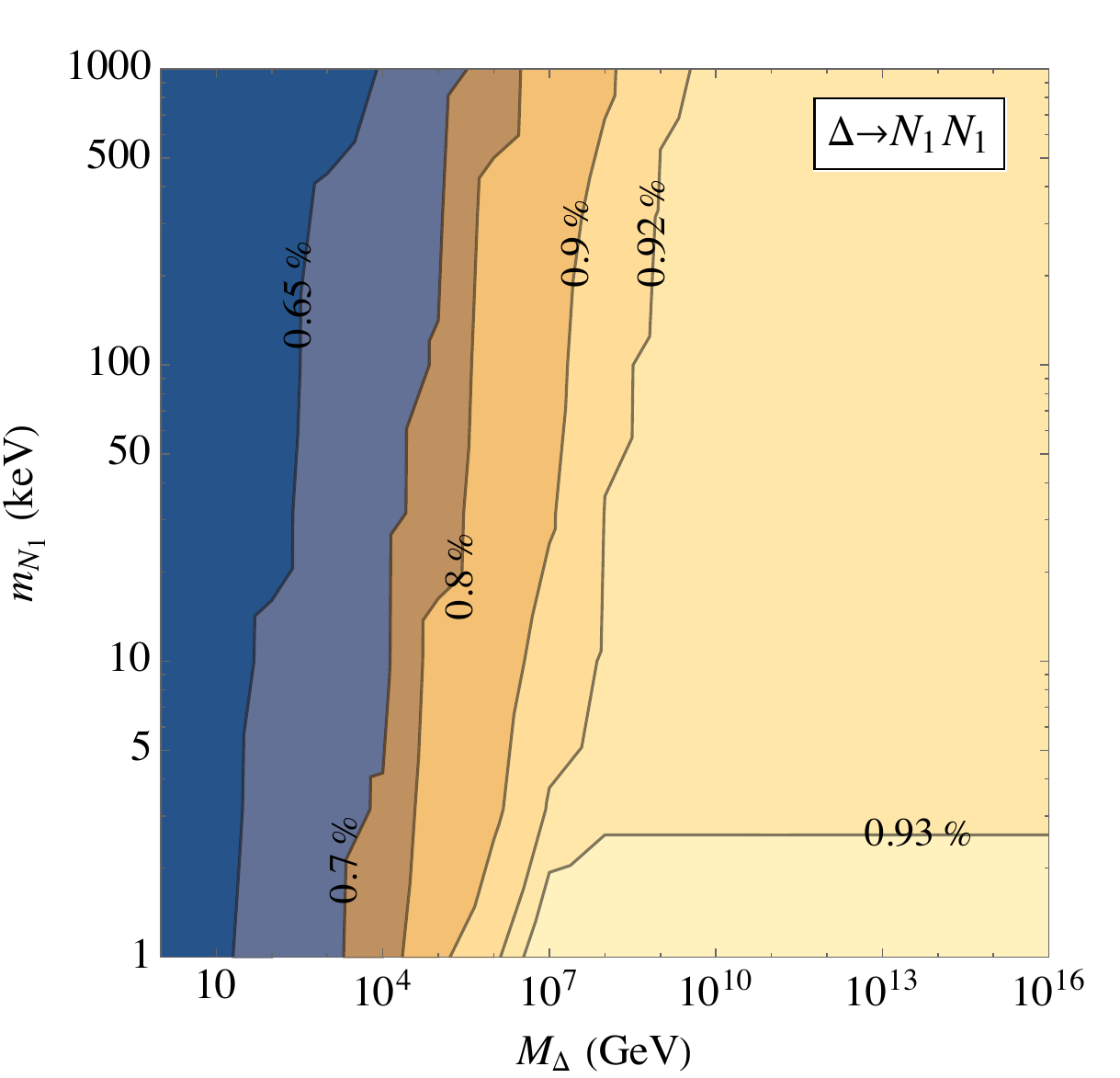}
  \end{center}
  \caption{
  Upper bound on $\text{Br}_X$, the branching ratio of dilutor $Y$ decaying into $X$ from the fit to LSS data
  (SDSS DR7 LRG), for the two dilution scenarios considered in TAB.~\ref{tab:models} in the context of LRSM. 
  For each point in the $m_X - m_Y$ parameter space, the $Y$ lifetime is determined by requiring $X$ to comprise 
  all of the dark matter in the universe.}
\label{fig:maxBr}
\end{figure}

The experimental data points come from the SDSS DR7 on luminous red galaxies~\cite{Reid:2009xm} (blue) and 
the Lyman-$\alpha$ forest~\cite{Lee:2012xb} (orange) measurements. 
All the curves in scenarios with secondary $X$ share a common feature with significant deviations from data 
in the $k \gtrsim 0.03\, h/{\rm Mpc}$ region. 
These occur at a much lower $k$ compared to other dark matter production mechanisms such as thermal 
freeze-in~\cite{DEramo:2020gpr, Decant:2021mhj}. 
This is mainly due to the large hierarchy between the dilutor and the dark matter mass, required by Eqs.~\eqref{app9}. 
Based on a simple $\Delta \chi^2$ fit to the data, we find that the LSS data from SDSS sets a much stronger 
constraint on these scenarios than Lyman-$\alpha$, making this probe particularly robust. 
The conflict with data increases with ${\rm Br}_X$, which translates into an upper bound, shown in FIG.~\ref{fig:maxBr} 
for the two models in TAB.~\ref{tab:models}. 
The bound does not depend much on the precise shape of the phase space distribution (because it is integrated
over) and the message is similar for both cases: the branching ratio of the dilutor decaying into dark matter is 
constrained to be 
\begin{equation}\label{eq:maxBr}
  \text{Br}_X \lesssim 1 \% \, , \quad @ 95\% \text{ CL} \, .
\end{equation}

This bound is nearly independent of $m_Y$, simply because the secondary component of the phase space 
distribution $f_X$ in~\eqref{eq:fX} is mostly independent of $m_Y$, as explained in the paragraph 
below~\eqref{eq:fX}.
The $\text{Br}_X$ limit gets slightly relaxed for larger $m_Y$, because holding the dark matter relic density fixed 
in Eq.~\eqref{app9} requires the lifetime $\tau_Y$ to be shorter, leading to a higher reheating temperature 
after the decay of dilutor. 
The corresponding temperature for the secondary dark matter component to become non-relativistic also increases, which 
is a $\sqrt[12]{g_*}$ effect, see Eq.~\eqref{eq:TNR}. 
Eventually, this shifts the deviation of $P(k)$ to a slightly higher $k$, where the data is less precise 
and thus becomes less constraining.

The constraint derived here comes predominantly from the LSS data, which relies only on the evolution of matter 
density perturbations in the linear regime. 
LSS thus provides a robust test of these models and we expect similar constraints to apply broadly for other dilutor $\to$ dark matter decay topologies.

Our result can be generalized to initial abundances for $Y$ and $X$ beyond the relativistic freeze-out. 
A sub-thermal initial population of $Y$ needs to be heavier and/or longer lived in order to provide the same amount 
of entropy injection.
The secondary $X$ particles from $Y$ decay become more energetic and take even longer to become matter-like. 
This impacts the primordial matter power spectrum down to even lower $k$ and leads to a more stringent constraint 
on ${\rm Br}_X$ than Eq.~\eqref{eq:maxBr}.
On the contrary, starting with a smaller overpopulation of $X$, the constraint on ${\rm Br}_X$ will be weaker.

%
%
\section{Anatomy of dilution scenarios in LRSM}\label{sec:theAnatomy}

Remarkably, the minimal LRSM contains all the ingredients for the dark matter dilution mechanism, described 
in the previous section, to occur. 
Throughout this work, we consider the lightest right-handed neutrino $N_1$ to be the dark matter, i.e., 
\begin{equation}
  X = N_1 \, . 
\end{equation}
Following the discussion in section~\ref{sec:N1DM}, with a mass below 100 MeV, it always freezes out relativistically 
and is generically overproduced.
The role of the diluting particle $Y$ can be played by either a heavier right-handed neutrino or the Majorana Higgs 
boson $\Delta$; the latter option is explored here for the first time.
In this section, we carefully examine the viability of each scenario and point out the corresponding
opportunities for experimental tests.

%
\subsection{Heavy neutrino dilutor} \label{sec:N2dilutor}

In this subsection, we first consider one of the two heavier right-handed neutrinos $N_2$ to play the role 
of the dilutor $Y$. 
We begin with the implications of LSS constraint obtained from the previous section, which results in
a non-trivial no-go theorem.
We then discuss the options for bypassing such constraints and describe the viable scenarios.

\subsubsection{Implication of the SDSS constraint}\label{sec:sdssLR}

Let us label the dilutor of section~\ref{sec:N2dilutor} as $Y = N_2$. 
The most obvious decay channels of $N_2$ are those mediated by the heavy $W_R$ boson in the LRSM. 
Corresponding Feynman diagrams for the decays are depicted in FIG.~\ref{fig:N2decayWR} below. 
Similarly to weak decays of the $\tau$ lepton in the SM, there are semi-leptonic and pure-leptonic decay channels.

\begin{figure}[ht]
  \begin{center}
    \includegraphics[width=0.618\textwidth]{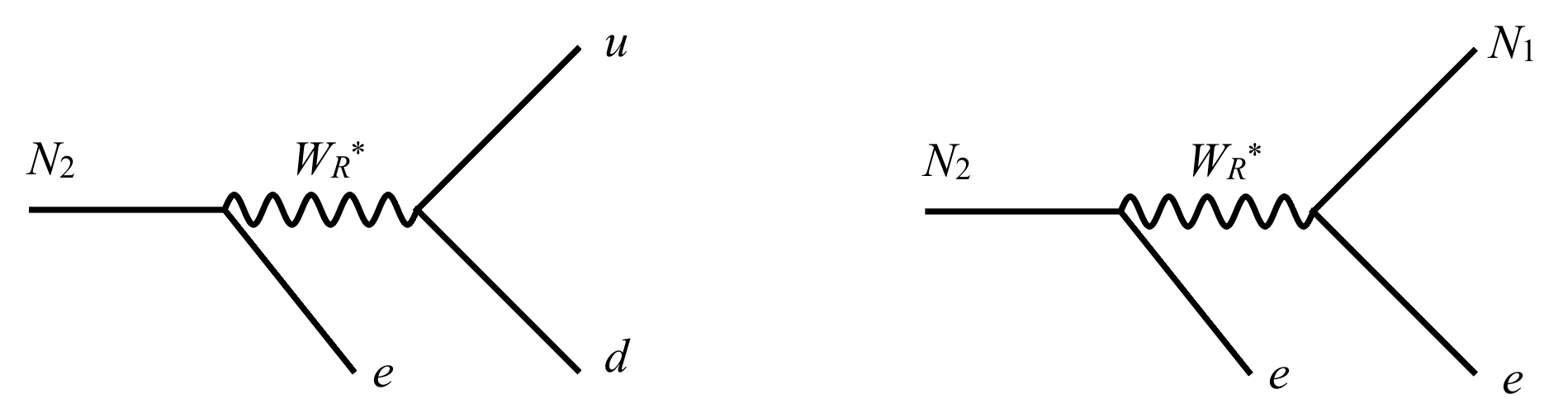}\vspace{-0.7cm}
  \end{center}
  \caption{
  Feynman diagram for dilutor $N_2$ decay via right-handed charged-current interaction mediated by $W_R$. 
  The flavor or final state charged leptons and quarks is dictated by the matrix elements of $V^R_{\rm PMNS}$ 
  and $V^R_{\rm CKM}$.}
  \label{fig:N2decayWR}
\end{figure}

All the decay products in the first channel are SM particles, which makes it a perfect decay mode for the 
DM dilution mechanism to work. 
In contrast, the second channel has a dark matter $N_1$ in the final state. 
The corresponding decay branching ratio is tightly constrained by large scale structure observations, to be smaller 
than 1\%, as pointed out in section~\ref{sec:LSS}. 

In the minimal LRSM, the ratios among the above $W_R$ mediated decay are fixed by the structure (particle content 
and gauge interactions) of the model.
If these were the only $N_2$ decay channels, the decay branching into the final state containing $N_1$ would be 
10\% for mass of $N_2$ well above the electroweak scale.
This branching ratio only gets higher for lighter $N_2$.
The large scale structure constraint from SDSS thus firmly excludes such simplest dilution scenario.
What may save the day are other possible decay channels of $N_2$. 
These may arise either due to neutrino or gauge boson mixing in the LRSM, as will be discussed next.

\subsubsection{A no-go theorem}\label{sec:nogo}

If minimality is one's first priority, it would be most desirable to open up the extra decay channel(s) for the 
right-handed neutrino dilutor and also allow it to participate in the type-I seesaw mechanism for generating the active 
neutrino masses. 
Na\"ively, both could occur through a mixing of $N_2$ with the active neutrinos. 
However, there is a no-go theorem against such a possibility.

If the dilutor $N_2$ participates in the seesaw mechanism, the mixing between $N_2$ and light neutrinos is bounded 
from below
\begin{equation}\label{eq:HLnuMixing}
  \theta_{N_2 \nu} \gtrsim \sqrt{\frac{m_\nu}{m_{N_2}}} \, .
\end{equation}
The value of $\theta_{N_2\nu}$ could be much larger than $\sqrt{{m_\nu^\odot}/{m_{N_2}}}$ due to additional degrees 
of freedom involved in type-I seesaw mechanism~\cite{Casas:2001sr}, but cannot be made smaller if $N_2$ is responsible
for neutrino mass generation.
For this contribution to the neutrino mass to be significant and at least explain the mass difference for solar 
neutrino oscillation, we require $m_\nu \gtrsim m_\nu^\odot \simeq \sqrt{8 \times 10^{-5} \,{\rm eV}^2}$. 

We first consider the case where the $N_2$ mass lies below the weak scale.
The partial decay width of $N_2$ via this mixing and an off-shell $W$ boson is then
\begin{equation}\label{eq:LifeN2}
  \tau_{N_2}^{-1} \gtrsim \frac{\mathfrak{m\,} G_F^2 m_{N_2}^5 \theta_{N_2\nu}^2}{96 \pi^3} 
  \gtrsim 10^{-23}\,{\rm GeV} \left( \frac{m_{N_2}}{1\,\rm GeV} \right)^4 \, ,
\end{equation}
where $\mathfrak{m}$ is the final state multiplicity factor within the range $\sim 1 - 10$.
In the first step, we neglect contributions to the decay via the $Z$ boson or possible interference effects.
This approximation would only affect our estimate by an order one factor, but keep our conclusion intact.
The first inequality also accounts for other possible (subdominant) $N_2$ decay channels (e.g., via $W_R$) and 
the second inequality follows from Eq.~\eqref{eq:HLnuMixing}. 
Plugging Eq.~\eqref{eq:LifeN2} into Eq.~\eqref{eq:RelicDilute}, we find a lower bound on the dark matter relic density
\begin{equation}
  \Omega_{N_1} \gtrsim 2.9 \left( \frac{m_{N_1}}{6.5 \, \rm keV} \right) \left(\frac{m_{N_2}}{1 \, \rm GeV} \right) \, .
\end{equation}
The reference mass for $N_1$ is the lower bound on warm dark matter found by the DES collaboration~\cite{DES:2020fxi}, 
which is consistent with other constraints using the Lyman-$\alpha$ forest~\cite{Viel:2013fqw, Irsic:2017ixq}, 
strong gravitational lensing observations~\cite{Gilman:2019nap, Hsueh:2019ynk}, and recent combined analysis~\cite{Enzi:2020ieg, Nadler:2021dft}.
The observed dark matter relic abundance then sets an upper bound on the mass of $N_2$,
\begin{equation}
  m_{N_2} \lesssim 90\,\rm MeV \, .
\end{equation}
Applying this bound again back in Eq.~\eqref{eq:RelicDilute}, we obtain a lower bound on the lifetime of $N_2$,
\begin{equation}\label{eq:conflict1}
  \tau_{N_2} \gtrsim 160 \,\rm sec \, .
\end{equation}
Because the universe was matter-dominated before the $N_2$ decayed away, the above lower bound on its lifetime is 
strongly excluded by the big-bang nucleosynthesis, which would require $\tau_{N_2} \lesssim 1 \text{ sec}$.

On the other hand, if $N_2$ is heavier than the weak scale, the decay induced by the mixing parameter 
Eq.~\eqref{eq:HLnuMixing} would be into an on-shell $W$ boson at a much higher rate,
\begin{equation}\label{eq:LifeN2b}
  \tau_{N_2}^{-1} \gtrsim \frac{G_F m_{N_2}^3 \theta_{N_2\nu}^2}{4\sqrt{2}\pi} \left(1 - \frac{M_W^2}{m_{N_2}^2} \right)
  \left(1 + \frac{M_W^2}{m_{N_2}^2} - \frac{2M_W^4}{m_{N_2}^4} \right) 
  \gtrsim 10^{-12} \, {\rm GeV} \left( \frac{m_{N_2}}{100\,\rm GeV} \right)^2 \, .
\end{equation}
The factor $G_F m_{N_2}^2$ properly accounts for a longitudinal enhancement in the limit when 
$m_{N_2}\gg M_W$.
Plugging this into Eq.~\eqref{eq:RelicDilute} leads to a lower bound on the dark matter relic density
\begin{equation}\label{eq:conflict2}
  \Omega_{N_1} \gtrsim 1.7\times10^5 \left( \frac{m_{N_1}}{6.5\,\rm keV} \right) \, .
\end{equation}
All the $N_2$ mass dependency cancels out completely and dark matter is considerably overproduced.

This completes the proof of the no-go theorem. 
It is derived by combining the constraints on the dark matter relic density and the dilutor lifetime.
Although we have used some $\sim$ in the above reasoning, the results in Eqs.~\eqref{eq:conflict1} 
and \eqref{eq:conflict2} are in sharp contradiction with the existing constraints, making it convincing 
there is no room to avoid the theorem.
It is also worth pointing out that the theorem not only applies to the LRSM focused in this work, but also 
to other gauge extensions, such as the $U(1)_{B-L}$ model~\cite{Davidson:1978pm, Mohapatra:1980qe, Bezrukov:2009th}.

\subsubsection{
\texorpdfstring{Type-II seesaw dominance and dilutor decay via $N-\nu$ mixing}{
Type-II seesaw dominance and dilutor decay via N-nu mixing}}

\begin{figure}[t]
  \begin{center}
    \includegraphics[width=0.7\textwidth]{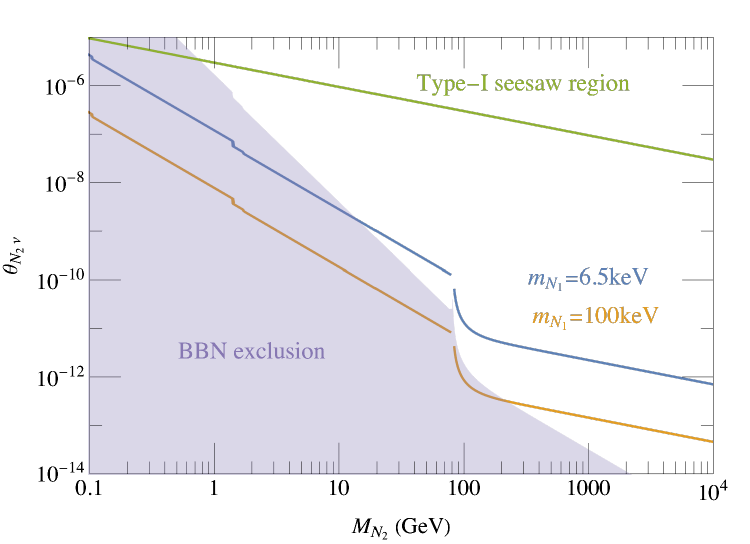} 
  \end{center}
  \caption{
  Parameter space for correct dark matter relic density, where $N_2$ serves as the dilutor and decays via a mixing 
  with the active neutrino. 
  We obtain $\Omega_{N_1} = 0.26$ along the blue and orange curves for $m_{N_1} = 6.5$ and 100\, keV, respectively.
  The purple shaded region is excluded by BBN, because the lifetime of $N_2$ is longer than a second.
  Along with the relic curves, it sets an upper bound on the mixing $\theta_{N_2\nu}$. 
  This upper bound is stronger for heavier $N_1$. 
  In contrast, the region above the green line in the upper-right corner of the figure shows the mixing angle needed for $N_2$ 
  to participate in the type-I seesaw mechanism and explain the neutrino mass difference for solar neutrino oscillation.
  This is clearly incompatible with the dilution mechanism and verifies the no-go theorem presented in section~\ref{sec:nogo}.
  } \label{fig:OnlyThetaNoXi}
\end{figure}

The no-go theorem presented above implies that if one of the heavier right-handed neutrinos (e.g., $N_2$) plays the 
role of dilutor, its mixing with light active neutrinos must be much smaller than Eq.~\eqref{eq:HLnuMixing}.
In other words, the contribution to neutrino mass from $N_2$ via the type-I seesaw must be well below what is 
needed for explaining the observed neutrino oscillation phenomena. 
At the same time, the dark matter candidate $N_1$ also cannot fully participate in the seesaw, because of the 
X-ray constraints.
The only remaining right-handed neutrino $N_3$ is free from constraints and does contribute to neutrino masses
in the type-I seesaw, but is unable to explain the two mass square differences needed for neutrino oscillations.
As a consequence, light neutrino masses must be accounted for by additional sources~\footnote{This argument also implies that the minimal $U(1)_{\rm B-L}$ model where the new gauge symmetry is broken by a Standard Model singlet scalar is unable to account for both the dark matter dilution mechanism and neutrino masses. Additional degrees of freedom (e.g. the couterpart of $\Delta_L$, see below) are needed.}.

A way out within the minimal LRSM is by considering another source of mass for the light neutrinos, which comes 
from the vacuum condensate of the left-handed scalar triplet $\Delta_L$, through the type-II seesaw mechanism.
By relieving the dilutor $N_2$ from the role of neutrino mass generation, we can treat its mixing with light 
neutrinos $\theta_{N_2\nu}$ as a free parameter, which can be arbitrarily small.
This enables a viable window in the model parameter space for the dark matter dilution mechanism to work.

In FIG~\ref{fig:OnlyThetaNoXi}, the blue and orange curves show where in the $\theta_{N_2\nu}$ versus $m_{N_2}$ plane 
the dark matter $N_1$ can obtain the correct relic density after the $N_2$ dilution, for two choices of $N_1$ mass.
The value 6.5\,keV is the lowest allowed warm dark matter mass by the DES result.
The purple region is excluded by BBN because the lifetime of $N_2$ is longer than a second. 
Clearly, viable values of $\theta_{N_2\nu}$ must be very tiny $\lesssim10^{-9}$ to satisfy both constraints.
In contrast, the region above the green line indicates the required values of $\theta_{N_2\nu}$ if $N_2$ participates in the type-I 
seesaw mechanism, which offers a way to visualize and quantigy the above no-go theorem.
We also find a lower bound on the dilutor $N_2$ mass of around 20 GeV.

The dark matter relic curves in FIG.~\ref{fig:OnlyThetaNoXi} are valid under the assumption that the $W_R$ mediated 
decay modes of the dilutor $N_2$ (see FIG.~\ref{fig:N2decayWR}) are subdominant to those induced by the $N_2 - \nu$ mixing. 
This condition is mostly easily satisfied if $N_2$ is heavier than the $W$ boson but still close to the weak scale. 
This leads to a lower bound on the mass scale of $W_R$ boson,
\begin{equation}\label{eq:WRlowerbound1}
  M_{W_R} \gtrsim \frac{10\, \rm GeV}{\sqrt{\theta_{N_2\nu}}} \gtrsim 10^6\,{\rm GeV} = 1\,{\rm PeV} \ ,
\end{equation}
where in the second step we read from FIG.~\ref{fig:OnlyThetaNoXi} that for $m_{N_2}>M_W$, the highest value of 
$\theta_{N_2\nu}$ is around $10^{-10}$, which is also orders of magnitude below the seesaw line, required for neutrino
mass generation.

\subsubsection{\texorpdfstring{Dilutor decay via $W-W_R$ mixing and $X$-ray limits}{
Dilutor decay via WL-WR mixing and X-ray limits}}\label{sec:type2Xray}

The previous subsections explored the possibility of the dilutor $N_2$ decaying dominantly through its mixing with the 
active light neutrinos.
Here, we discuss another possible $N_2$ decaying channel inherent to the minimal LRSM, via the gauge boson mixing 
$\xi_\text{LR}$, mentioned in section \ref{sec:WWRmix} and shown in the Feynman diagram on FIG.~\ref{fig:N2decayW} below.
Because the absolute value of $\xi_\text{LR}$ is bounded from above by $M_W^2/M_{W_R}^2$ (see Eq.~\eqref{eq:xiLR}), if $N_2$ 
is lighter than the $W$ boson and decays via off-shell $W$, the corresponding partial decay rate will have the same 
parametric dependence as those in Fig.~\ref{fig:N2decayWR}. 
It cannot provide sufficient suppression to the branching ratio of $N_2\to N_1$ decay and the resulting dark matter 
production/dilution mechanism still suffers from the strong constraint from large scale structure. 
This observation forces the viable parameter space to the window where $M_{W_R} > m_{N_2} > M_W$. 

\begin{figure}[ht]
  \begin{center}
    \includegraphics[width=0.25\textwidth]{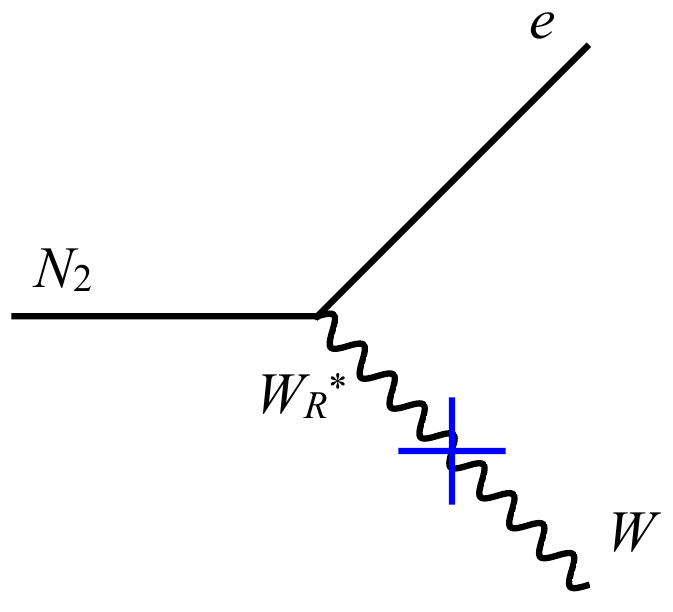}\vspace{-0.7cm}
  \end{center}
  \caption{
  Feynman diagram for $N_2$ decay via $W-W_R$ mixing in LRSM. 
  Blue cross indicates an insertion of $\xi_\text{LR}$ mixing.}
  \label{fig:N2decayW}
\end{figure}

In this case, the available decay rates for $N_2$ are
\begin{equation}\label{eq:xiLRN2br}
\begin{split}
 \Gamma_{N_2\to N_1 \ell^+\ell'^-} &= \frac{G_F^2 m_{N_2}^5}{96\pi^3} \left( \frac{M_W}{M_{W_R}} \right)^4 \, , \qquad
 \Gamma_{N_2\to \ell q \bar q'} = \frac{\mathfrak{m\,} G_F^2 m_{N_2}^5}{96\pi^3} \left( \frac{M_W}{M_{W_R}} \right)^4 \, , 
 \\
 \Gamma_{N_2\to \ell W} &= \frac{g^2 |\xi_\text{LR}|^2 m_{N_2}}{32\pi} \left(\frac{m_{N_2}}{M_W}\right)^2 
 \left(1 - \frac{M_W^2}{m_{N_2}^2}\right) \left(1 + \frac{M_W^2}{m_{N_2}^2}- \frac{2M_W^4}{m_{N_2}^4}\right) \, , 
\end{split}
\end{equation}
where the first two decays occur via diagrams in FIG.~\ref{fig:N2decayWR} and the last one via FIG.~\ref{fig:N2decayW}.
Again, $\mathfrak{m}$ is the final state multiplicity factor, which equals 12 (9) for $m_{N_2}>(<)m_t+m_b$.
In the absence of $\theta_{N_2\nu}$, the three rates in~\eqref{eq:xiLRN2br} sum up to the total decay rate of $N_2$.

\begin{figure}[t]
  \begin{center}
    \includegraphics[width=0.45\textwidth]{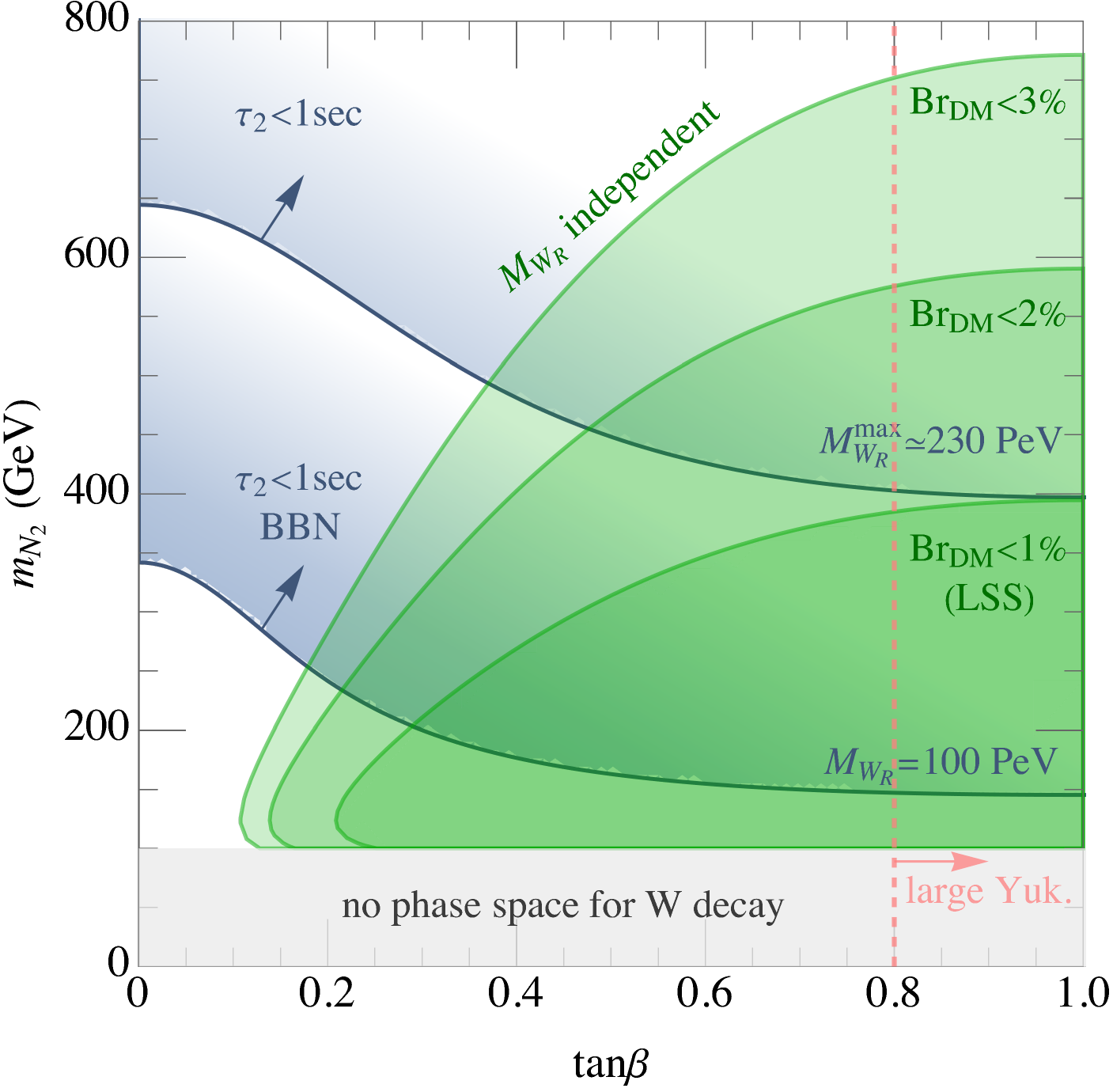} 
    \includegraphics[width=0.5\textwidth]{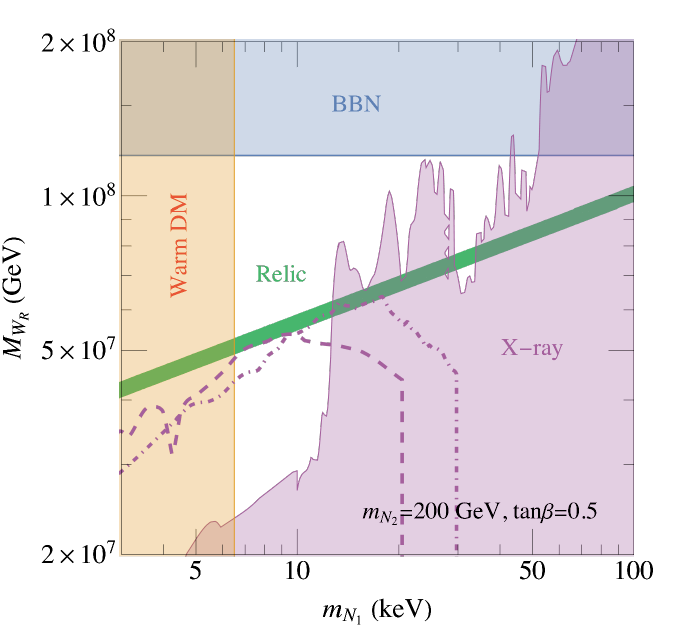}
  \end{center}
  \caption{
  %
  Left: Viable parameter space for the dark matter dilution to work in the $m_{N_2}$ versus $\tan\beta$ plane, where 
  the diluting particle $N_2$ dominantly decays via the mixing between $W$ and $W_R$ gauge bosons. 
  The region outside the darkest green is excluded by the LSS constraint in Eq.~\eqref{eq:maxBr}. 
  The region to the right of the vertical pined dashed line is excluded by the theoretical constraint on the range of $\tan\beta$, 
  Eq.~\eqref{eq:tanbetarange}.
  The two dark blue curves corresponds to lower bound on $m_{N_2}$ to pass the BBN constraint, for two 
  choices of $W_R$ mass, $10^8$\,GeV and $2.25\times 10^8$\,GeV, respectively. 
  %
  %
  Right: Further constraints in the $M_{W_R}$ versus $m_{N_1}$ plane, with other parameters fixed, $m_{N_2}=200\,$GeV 
  and $\tan\beta = 0.5$. 
  The blue shaded region again indicates the upper bound on $M_{W_R}$ from the BBN constraint. 
  The purple region is excluded by the existing $X$-ray line searches for dark matter decay $N_1\to \nu \gamma$ via the 
  loop processes shown in Fig.~\ref{fig:N1decayW}. 
  The dashed and dot-dahsed purple curves corresponds to the reach of future $X$-ray experiments ATHENA
  and XRISM, respectively.
  The orange shaded region is the warm dark matter exclusion limit set by the DES experiment. 
  The green band is where dark matter obtains the correct relic density after the entropy dilution.
   }\label{fig:type2}
\end{figure}

Like before, we impose three requirements on the dilution scenario here:
\begin{enumerate}
    \item Dark matter $N_1$ obtains the correct relic density;
    \item The decay branching ratio of dilutor $N_2$ to dark matter is smaller than 1\%;
    \item Dilutor decays faster than 1 second.
\end{enumerate}
Our main results are then summarized in FIG.~\ref{fig:type2}. 

The left-panel shows the parameter space of $m_{N_2}$ versus $\tan\beta$. 
Outside the darkest green shaded region, the dilutor to dark matter decay branching ratio exceeds 1\% and the parameter 
space is excluded by the LSS data. 
From Eq.~\eqref{eq:xiLRN2br}, it is useful to note that the branching ratio and the LSS constraint is independent of other 
parameters of the model such as $M_{W_R}$ or $m_{N_1}$.
In contrast, the total decay rate of dilutor $N_2$ does depends on $M_{W_R}$ and so is the allowed parameter space that 
is consistent with the BBN constraint. 
The two blue curves show the lower limit on $m_{N_2}$ for two choices of $M_{W_R}=10^8$\,GeV (lower) and $2.25\times10^8\,$GeV
(upper), respectively. 
Clearly, the latter case is marginal where the available parameter space for dilution mechanism closes. 
From this, we derive a upper bound on $M_{W_R}<2.25\times10^8$\,GeV.
In the same plot, the vertical pink dashed line corresponds to the theoretical upper bound on $\tan\beta$ derived in 
Eq.~\eqref{eq:tanbetarange}.
Viable parameter space for dark matter relic only occurs in the darkest green region with proper arrangement of other parameters.

Next, we address the $X$-ray line search limits on dark matter $N_1$ decay, as shown on the right panel of FIG.~\ref{fig:type2}.
In the minimal LRSM, the dark matter candidate is not absolutely stable and there are in fact two contributions to the radiative decay 
of $N_1 \to \nu \gamma$. 
One is via the $N_1$-$\nu$ mixing and applies also to the regular sterile neutrino dark matter. 
The other is through the $W-W_R$ mixing and both occur at the one loop level with coherent amplitudes.
In the presence of a nonzero $\theta_{N_1 \nu}$, the radiative decay rate of dark matter is a well known result~\cite{Shrock:1974nd, Pal:1981rm},
\begin{equation}
  \Gamma_{N_1 \to \nu \gamma} = \frac{9 \alpha \xi_\text{LR}^2}{256 \pi^4} G_F^2 m_{N_1}^5 \sin^2\theta_{N_1\nu} \, .
\end{equation}
In the presence of a nonzero $\xi_\text{LR}$, there are new Feynman diagrams for the radiative decay of dark matter $N_1$. 
%
We derive the leading-order decay rate
\begin{equation}\label{eq:N_1radiativexiLR}
  \Gamma_{N_1\to \nu\gamma} = \frac{\alpha \xi_\text{LR}^2}{8\pi^4} G_F^2 m_{N_1}^3 
  \sum_\ell \left| \left( V^R_{\rm PMNS} \right)_{\ell1} \right|^2 m_\ell^2 \, .
\end{equation}
See appendix~\ref{appdx1} for a detailed derivation of this rate.
In this case, because of the $W_R$ and $\tan \beta$ dependencies in $\xi_\text{LR}$, we find a closer interplay between the 
$X$-ray search bounds and the requirements on the dilution mechanism found in the previous subsection.

The implications from $X$-ray constraints are showns in FIG.~\ref{fig:type2} (right).
We fix $m_{N_2} = 200\,$GeV and $\tan\beta = 0.5$ which is an allowed point in FIG.~\ref{fig:type2} (left), and show the 
other constraint in the dark matter mass $m_{N_1}$ versus $M_{W_R}$ parameter space.
As discussed earlier, for the dilutor $N_2$ to decay before BBN, there is an upper bound on $M_{W_R}$ for given $m_{N_2}$. 
This excludes the blue shaded region. 
The purple shaded region is then excluded by the existing $X$-ray line searches for dark matter decay~\cite{Boyarsky:2007ge, Merle:2013ibc, Ng:2019gch}, 
which sets a lower bound on $M_{W_R}$ and upper bound on $m_{N_1}$. 
Interestingly, the remaining window for viable dark matter in this scenario can be tested by the upcoming X-ray 
experiments ATHENA~\cite{Neronov:2015kca} 
  and XRISM~\cite{XRISMScienceTeam:2020rvx, Dessert:2023vyl}, as shown by the dashed and dot-dashed purple curves.
Here, we assume generic flavor mixing matrix where all the elements are $\mathcal{O}(1)$ in magnitude and approximate $\sum_\ell |(V^R_{\rm PMNS})_{\ell 1}|^2 m_\ell^2 \sim m_\tau^2$ in 
Eq.~\eqref{eq:N_1radiativexiLR}.
The $X$-ray limit could be weakened if the right-handed leptonic mixing matrix element $(V^R_{\rm PMNS})_{\tau1}$ is suppressed
or if destructive interference between amplitudes is strong enough.
As explained in section~\ref{sec:psd} (see also Eq.~\eqref{eq:fX}), for sufficiently small ${\rm Br}_{Y\to X}$, the phase space 
distribution of dark matter $N_1$ follows exactly the thermal distribution, exactly like a warm dark matter is 
defined~\cite{Colombi:1995ze, Viel:2005qj}. 
The orange shaded region corresponds to a lower bound of 6.5\,keV on warm dark matter mass.

\subsubsection{
\texorpdfstring{Lower bound on the $W_R$ mass scale}{Lower bound on the WR mass scale}}

So far, we have discussed several options of having the light right-handed neutrino $N_1$ to comprise all the dark matter in 
the universe, through the dilution mechanism where the dilutor is a heavier right-handed neutrino $N_2$. 
In all cases, we find that the mass scale of the $W_R$ gauge boson must be rather high. 
In the case where $N_2$ dominantly decays via its mixing with light neutrinos, the lower bound is around PeV scale, as found in 
Eq.~\eqref{eq:WRlowerbound1}. 
In the case where $N_2$ dominantly decays via $W -W_R$ gauge boson mixing, lower bound on $M_{W_R}$ is higher (tens of PeV), due to the relic 
density explanation and constraints on the radiative decay of dark matter $N_1$, as shown in FIG.~\ref{fig:type2} (right).
In both cases when calculating the decay rate of dilutor $N_2$, we have made the assumption that the masses of its decay products 
are much smaller than $m_{N_2}$.
The lower bounds on $M_{W_R}$ are derived based on this assumption, which are generic and does require special arrangement of the 
parameters of the model.

Here, we wish to scrutinize if the mass scale of $W_R$ is allowed to be even lighter at all if some amount of tuning of parameters
is arranged. 
While this might be less appealing, the main motivation behind is the prospect of other experimental probes (such as high-energy 
colliders) of the LR symmetry scale.
Such a possibility was first explored in~\cite{Nemevsek:2012cd}, which resorts to a compress spectrum with the mass of dilutor $N_2$ being 
close to the sum of charged pion and a charged lepton masses. 
This leads to a phase space suppression in the dilutor decay rate ($N_2\to \pi + \ell$) and enables sufficient longevity while keeping 
$W_R$ mass near the TeV scale.
Moreover, the flavor structure of the right-handed lepton mixing matrix $V^R_{\rm PMNS}$ must also be tuned, such that $N_1$ primarily
couples to the $\tau$ lepton in the right-handed current interaction, thus kinematically forbidding the $N_2\to N_1$ decay (via off-shell 
$W_R$) that is constrained by large scale structure as discussed in section~\ref{sec:sdssLR}. 
In this scenario, light neutrino masses can be explained via a mixed type-I (where $N_3$ mainly contributes) and type-II seesaw mechanism.

To explain the dark matter relic density using the dilution mechanism (see Eq.~\eqref{eq:RelicDilute}), it seems challenging to have a dilutor 
$N_2$ mass well below 2.2 GeV, because BBN forbids the lifetime of $N_2$ to be longer than a second, while at the same time the Tremaine-Gunn 
bound forbids dark matter mass to be well below a keV.
To get around this difficulty, \cite{Nemevsek:2012cd} noticed a special mass window around $M_{W_R}\sim 5$ TeV could work, where the above flavor structure 
allows dark matter $N_1$ to freeze out slightly before the QCD phase transition whereas the dilutor $N_2$ freezes out slightly after. 
It leads to an enhancement factor in the dilution factor $\mathcal{S}$ in Eq.~\eqref{eq:dilutionfactorS}, given by the ratio of $g_*$ at 
temperatures above and below $\Lambda_{\rm QCD}$, and in turn a suppression in the final dark matter relic density. 
Thanks to this effect, \cite{Nemevsek:2012cd} found viable solutions for dark matter mass around 0.5 keV. 
However, after the recent substantial progress in constraining the warm dark matter mass, e.g. $m_{N_1} > 6.5\,$keV, found by the 
DES collaboration from ultra-faint dwarfs~\cite{DES:2020fxi}, such a low mass $W_R$ window has been firmly closed.
This leads us to conclude that with a right-handed neutrino dilutor, the up-to-date lower bound on $M_{W_R}$ for 
consistent dark matter cosmology in LRSM is pushed to above the PeV scale, given by Eq.~\eqref{eq:WRlowerbound1}.

%
\subsection{Majorana Higgs dilutor}\label{sec:HiggsDilution}

In this subsection, we explore the other dilutor candidate in LRSM, the Majorana Higgs $\Delta$, introduced in section~\ref{sec:Delta}.
The role of dark matter is still played by $N_1$.
To our knowledge, such a possibility has not been considered in any previous dark matter analysis of the model.

\begin{figure}[ht]
  \begin{center}
    \includegraphics[width = 0.618\textwidth]{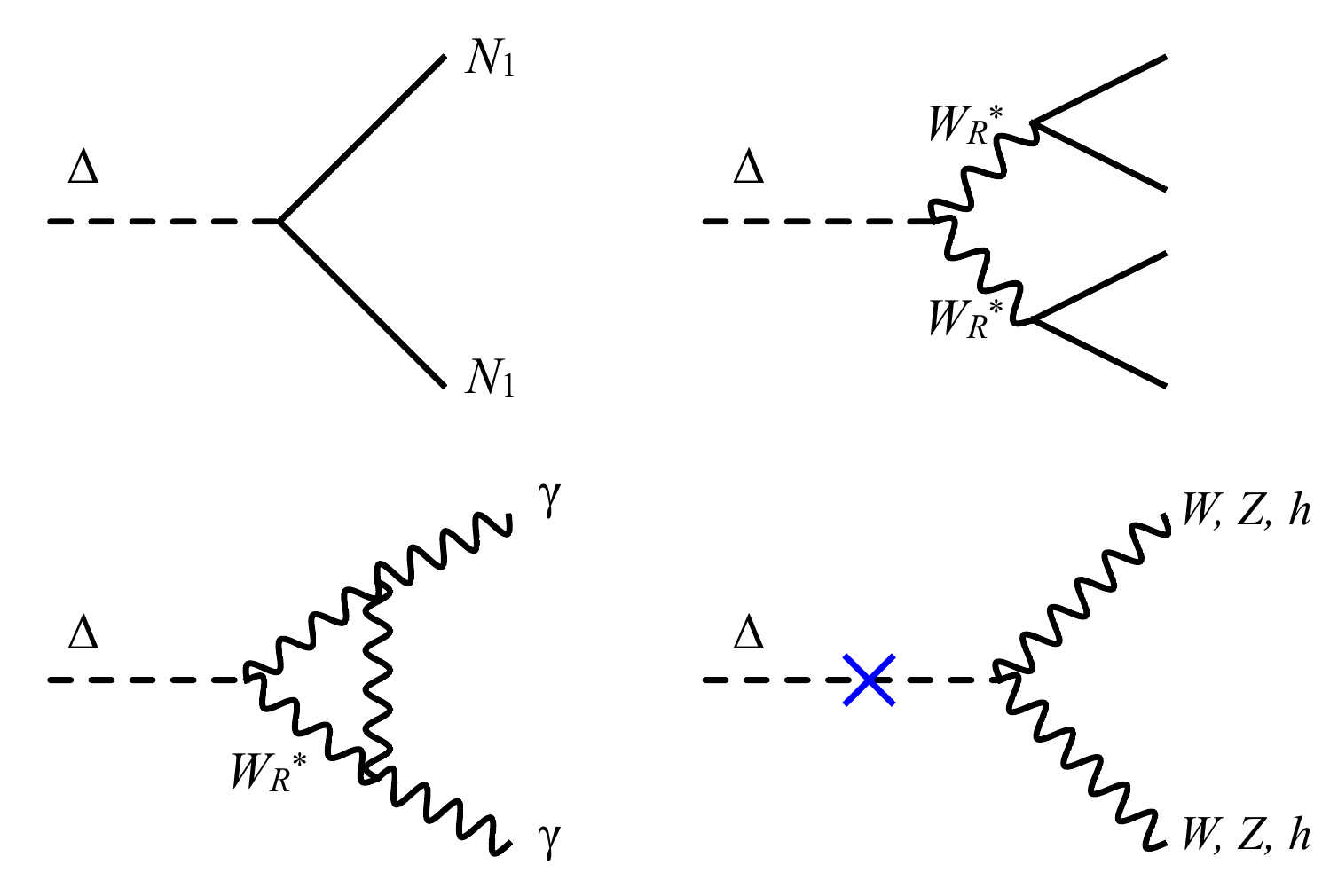}\vspace{-0.7cm}
  \end{center}
  \caption{
  Feynman diagrams for the decay rate of the dilutor $\Delta$ in the minimal LRSM. 
  The coupling of $\Delta$ to right-handed neutrino and $W_R^\pm$ are proportional to their masses, whereas the blue cross in 
  the last diagram represents $\Delta$-Higgs boson mixing, generated by the scalar potential. 
  In the second diagram, the fermion final states that connect to virtual $W_R$ include quark pairs and $N_1$ plus 
  a charged lepton. 
  }
  \label{fig:DeltaDecay}
\end{figure}

Through the spontaneous gauge symmetry breaking $SU(2)_R \times U(1)_{B-L} \to U(1)_Y$, the couplings of $\Delta$ are tied to mass 
generation of the gauge bosons $W_R^\pm, Z'$ and the right-handed neutrinos $N_i$.
To be long lived and qualify as the diluting particle, the mass of $\Delta$ must be well below those of $W_R$ and $Z'$. 
We will work in the parameter space where $\Delta$ is also much lighter than two of the right-handed neutrinos $N_{2}, N_3$. 
As a result, the decays of $\Delta$ into these on-shell final state particles are forbidden.
Its possible decay channels are shown by the Feynman diagrams in FIG.~\ref{fig:DeltaDecay}.

The decay rate of $\Delta$ into two dark matter particles $N_1$ is suppressed by the small mass $m_{N_1}$. 
The decays via off-shell $W_R$ into light fermions or two photons are suppressed by the small ratio of $M_\Delta/M_{W_R}$. 
The decay via off-shell $Z'$ is always subdominant, because $Z'$ is heavier than $W_R$ by a factor of $\sqrt{3}$ and a relatively 
smaller decay branching ratio into $N_1$~\cite{Mohapatra:1987gb}. 
Finally, $\Delta$ could decay via a mixing with the SM Higgs boson. 
For $M_\Delta$ well above the electroweak scale, the dominant decays via the Higgs mixing are into $W^+ W^-, Z Z$ and 
$h h$ final states.

With the mass hierarchy $M_W \ll M_\Delta \ll M_{W_R}$, the partial decay rates of $\Delta$ are
\begin{equation}\label{eq:Deltabr}
\begin{split}
  \Gamma_{\Delta\to N_1N_1} &= \frac{G_F M_W^2 m_{N_1}^2 M_\Delta}{4\sqrt{2}\pi M_{W_R}^2} \, , \qquad
  \Gamma_{\Delta\to W_R^*W_R^*} = \frac{5G_F^3 M_W^6 M_\Delta^7}{576\sqrt{2} \pi^3 M_{W_R}^6} \, , 
  \\
  \Gamma_{\Delta\to \gamma\gamma} &= \frac{49 \alpha^2 G_F M_W^2 M_\Delta^3}{128\sqrt{2}\pi^3 M_{W_R}^2} \, , \qquad
  \Gamma_{\Delta\to h^*} \simeq \frac{\theta_{\Delta h}^2 G_F M_\Delta^3}{4\sqrt{2} \pi} \, .
 \end{split}
\end{equation}
For simplicity, we work in the limit where all final state particle masses are negligible.
The $\Delta \to W_R^* W_R^*$ decay occurs through two off-shell $W_R^\pm$ bosons and has four right-handed fermions 
in the final states.
For this partial rate, we apply the 4-body decay formula Eq.~(2.35) of~\cite{Djouadi:2005gi} and work in the heavy $W_R$ limit. 
The kinematically allowed fermion final states are quark pairs and $N_1$ plus a charged lepton.

Among the above four decay channels of the dilutor, the first two can produce energetic dark matter $N_1$ in the final 
state and get constrained by the large scale structure observations.
The third channel, where $\Delta$ decays into a pair of photons, can bypass the LSS constraint.
Indeed, we find that the ratios
\begin{align}\label{eq:RatiosOfGamma}
 \frac{\Gamma_{\Delta\to N_1N_1}}{\Gamma_{\Delta\to \gamma\gamma}}&\simeq 1.2\times10^{-7} 
 \left(\frac{m_{N_1}}{M_\Delta}\right)^2 \ , 
 &
 \frac{\Gamma_{\Delta\to W_R^*W_R^*}}{\Gamma_{\Delta\to \gamma\gamma}} &\simeq 2.3 \times 
 \left(\frac{M_\Delta}{M_{W_R}}\right)^4 \ ,
\end{align}
can both be made much smaller than 1\% if $m_{N_1} \ll M_\Delta \ll M_{W_R}$. 
This mass-scale hierarchy is consistent with the above mass spectrum assumptions. 
It allows the LRSM to evade the LSS constraint, even in the absence of $\Delta-$Higgs boson mixing.

\begin{figure}[t]
  \begin{center}
    \includegraphics[width=0.48\textwidth]{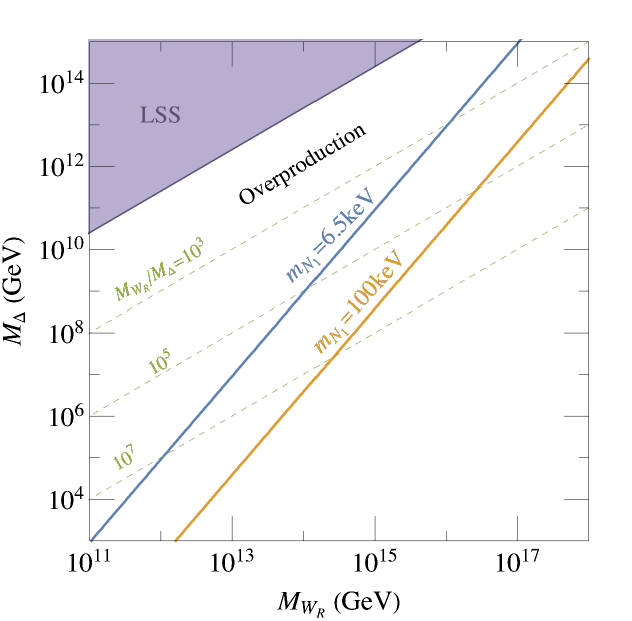}\quad \includegraphics[width=0.48\textwidth]{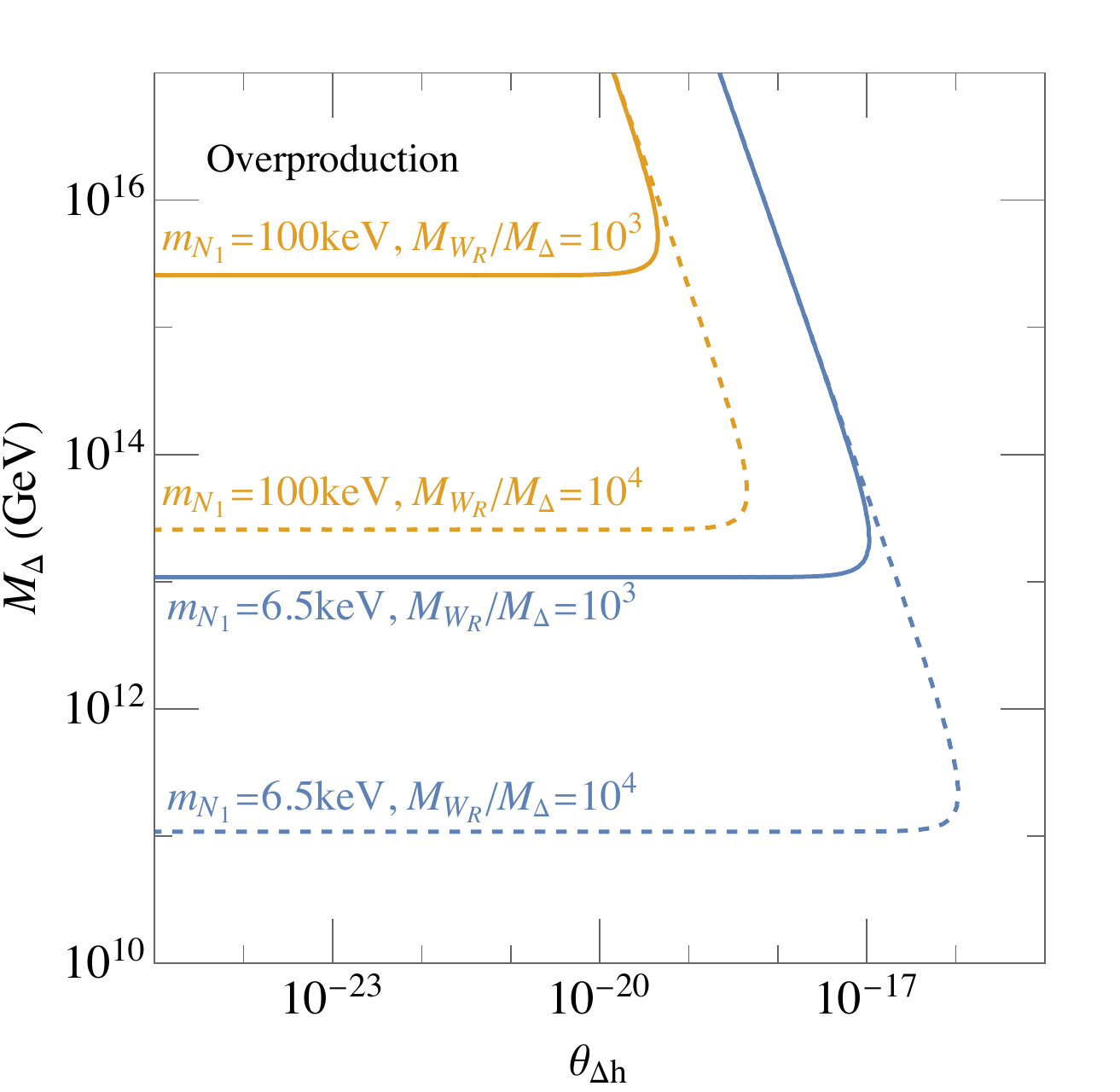}
  \end{center}
  \caption{
  Blue and orange curves show the parameter space where dark matter obtains the correct relic abundance via the dilution mechanism, 
  where the Majorana Higgs $\Delta$ plays the role of dilutor.  
  In the left panel, we set $\theta_{\Delta h} = 0$. 
  The purple region is excluded by the LSS observations. 
  The green dashed curves indicate different $M_{W_R}/M_\Delta$ mass ratios.
  In the right panel, we turn on $\theta_{\Delta h}$ but fix the $M_{W_R}/M_\Delta$ mass ratio for each relic curve.}
  \label{fig:Delta}
\end{figure}

In the left panel of FIG.~\ref{fig:Delta}, we first work in the $\theta_{\Delta h} = 0$ limit and the blue and orange curves 
show the $M_\Delta$ versus $M_{W_R}$ parameter space, where dark matter $N_1$ obtains the correct relic abundance through the $\Delta$-dilution
mechanism, for two values of $m_{N_1}=6.5\,$keV and 100\,keV, respectively. 
We apply Eq.~\eqref{eq:RelicDensityAfterDilution} by identifying $Y = \Delta$.
Dark matter is overproduced in regions to the left of the curves. 
The purple region is excluded by LSS observations, because the branching ratio for $\Delta\to W_R^*W_R^*\to \text{light fermions}$
decay is too high (see Eq.~\eqref{eq:RatiosOfGamma}). 
We find that the mass scale of $W_R$ must be very high, above $\sim10^{11}$\,GeV, but the mass scale of $\Delta$ can be much lower. 
However, the price of having a lighter $\Delta$ is to increase the mass hierarchy between $\Delta$ and $W_R$, as indicated by the 
green dashed lines. 
Similar to the argument in footnote~\ref{footVacuum} against a very light $\Delta$ to be the dark matter, we do not consider $\Delta$ 
to be lighter than $W_R$ by much more than a loop factor.
Taking into account of this theoretical constraint, we end up finding that both $\Delta$ and $W_R$ masses are pushed to rather 
high values, close to the GUT scale.
We have also checked that the lifetime of $\Delta$ is much shorter than 1 second along the entire curves, thus safely evading 
the BBN constraint.

In the right panel of FIG.~\ref{fig:Delta}, we project the relic curves to the $\theta_{\Delta h}$ versus $M_\Delta$ parameter space 
for two realistic choices of the mass ratio $M_{W_R}/M_\Delta$.
On each relic curve, the horizontal part has no $\theta_{\Delta h}$ dependence because the total decay rate is dominated 
by $\Delta \to \gamma \gamma$.
The decay via Higgs mixing takes over in the region with larger $\theta_{\Delta h}$ and heavier $\Delta$.
On each relic curve, there is also an upper bound on $\theta_{\Delta h}$ otherwise $\Delta$ would decay too fast.

To justify the use of Eq.~\eqref{eq:RelicDensityAfterDilution}, we must verify that $\Delta$ decouples from the rest of the plasma 
while it was still ultra-relativistic. 
First of all, at temperatures around the LR symmetry breaking scale, $\Delta$ is in thermal equilibrium with heavy particles that 
receive their mass from $v_\Delta$, which are $N_{2,3}$ and $W_R, Z'$.
When the temperature of the universe falls around the $M_\Delta$, the $N_{2,3}$ and $W_R, Z'$ particles already decouple from the 
thermal plasma because they are much heavier.
The remaining processes to consider are similar to those in FIG.~\ref{fig:DeltaDecay}.
Among them, the process $\Delta\leftrightarrow \gamma\gamma$ is suppressed by the heavy $W_R$ mass and remains decoupled until the temperature of the universe cools down to (set by $\Gamma_{\Delta\leftrightarrow\gamma\gamma}=H$)
\begin{equation}
  T \sim 10^9 \,{\rm GeV} \left(\frac{M_\Delta}{10^{13}\,\rm GeV} \right)^{1/2} \left( \frac{10^3}{M_{W_R}/M_\Delta} \right) \, .
\end{equation}
This temperature is well below $M_\Delta$, thus the inverse decay will be Boltzmann suppressed and never reach equilibrium. 
As discussed in Eq.~\eqref{eq:RatiosOfGamma}, the other processes $\Delta\leftrightarrow N_1 N_1$ and $\Delta \leftrightarrow 4q$ 
have rates much smaller than $\Delta\leftrightarrow \gamma\gamma$ and cannot keep $\Delta$ thermalized either.
Therefore, the decoupling of $\Delta$ must occur at a temperature between $M_{W_R}$ and $M_\Delta$.
The particle mass spectrum considered for the $\Delta$-dilution mechanism is indeed compatible with the assumption that $\Delta$ 
freezes out relativistically.

%
%
\section{Conclusion and Outlook}\label{sec:conclusion}

In this work, we explore the entropy dilution mechanism for dark matter relic density in the minimal LRSM that also addresses 
the origin of neutrino mass.
In this model, the lightest right-handed neutrino ($N_1$) is the sole dark matter candidate and its mass must be below the QCD 
scale in order to stay cosmologically stable.
We first emphasize that $N_1$ always decouples relativistically from the right-handed current interactions and an entropy release 
afterwards must happen for producing the observed dark matter relic abundance in the universe. 
This requires the presence of a ``long-lived'' diluting particle which comes to dominate the energy content of the universe as a
matter component, before decaying away mainly into SM particles. 
One of the heavier right-handed neutrinos ($N_2$) or the Higgs boson from spontaneous $SU(2)_R\times U(1)_{B-L}$ gauge symmetry 
breaking ($\Delta$) can play the role of the diluting particle.

Our original contribution here is a new opportunity of such a mechanism in cosmology. 
The matter power spectrum for the large scale structure of the universe is sensitive to the diluting particle's partial decay 
model into dark matter. 
When produced this way, dark matter can remain relativistic until the onset of recombination and suppress the primordial 
matter density perturbations. 
Through a detailed analysis, we derive an upper bound on such a decay branching ratio to be less than $\sim 1\%$, using the 
existing SDSS data. 
Such a large scale structure constraint is generic and can be applied to various dark matter models that require 
an entropy dilution mechanism. 
In the context of LRSM, the decay of $N_2$ into $N_1$ can happen via right-handed charged-current interaction (mediated by the 
$W_R^\pm$ gauge boson) and the decay of $\Delta$ into $N_1$ is tied to dark matter mass generation. 
Therefore, the large scale structure constraint plays a crucial role in determining the viable parameter space for the dark 
matter relic density. 
We carry out an anatomy of possible dark matter dilution scenarios in the left-right symmetric model:
\begin{itemize}
    \item In the scenario of $N_2$ dilution, we point out that the decays of $N_2$ cannot be dominated by the right-handed 
    currents, otherwise the $N_2\to N_1$ branching ratio is too high ($\gtrsim 10\%$).
    Thus, additional decay modes must be present due to a $N_2-$light-neutrino mixing or $W-W_R$ gauge boson mixing. 
    In both cases, we find that the mass scale of the $W_R$ boson must be rather high, above the PeV scale. 
    On the other hand, the dilutor $N_2$ can have a mass as low as the weak scale. 
    We also derive the mono-chromatic $X$-ray constraint on dark matter $N_1$ from $W-W_R$ mixing which further narrows down the 
    viable mass range of $N_1$ to a mass window of $6.5-30$ keV. 
        
    \item The possibility of $\Delta$ dilution is original to this work. 
    This scenario requires $\Delta$ to be lighter than the right-handed gauge bosons and neutrinos (except for $N_1$). 
    We find the $\Delta \to \gamma \gamma$ mode to be the most useful for suppressing the 
    $\Delta \to N_1 N_1$ decay and passing the strong large scale structure constraint. 
    The corresponding mass scales of $\Delta$ and $W_R$ need to be very high, close to the GUT scale.
    Because of this, the $W$-$W_R$ mixing contribution to $N_1$ dark matter radiative decay is negligible.
    The $N_1\to \nu\gamma$ decay will only proceed via its mixing with light active neutrinos, as is the case of a regular sterile neutrino.
\end{itemize}

Based on the above results, we point out the following opportunities of diluted dark matter in the light of the upcoming experimental efforts:
\begin{itemize}
    \item The primordial dark matter power spectrum will be more precisely measured by the upcoming large scale galaxy surveys, including 
    Euclid and Rubin LSST~\cite{Euclidtalk, Ferraro:2022cmj}.
    Future high-redshift surveys such as MegaMapper and PUMA have the promise to extend the precision measurement up to 
    wave number $k\sim 0.9\,h/{\rm Mpc}$~\cite{Sailer:2775916}. 
    A discovery of suppressed matter power spectrum, starting from $k\gtrsim 0.03\,h/{\rm Mpc}$, will serve as a smoking-gun evidence 
    for dark matter entropy dilution mechanism in the early universe with a nonzero dilutor to dark matter decay branching ratio.
    \item Related to the thermal dark matter population that gets diluted, future experimental facilities exploring the small scale 
    structure of the universe will be instrumental as well.
    Observations of low mass dark matter halos and the lensing of cosmic microwave background may allow the discovery and measurements 
    of the dark matter mass, if it lies not far above the current lower bound ($\sim6.5\,$keV)~\cite{Chakrabarti:2022cbu, Drlica-Wagner:2022lbd}.
    \item Future experiments including ATHENA and XRISM~\cite{Neronov:2015kca, XRISMScienceTeam:2020rvx, Dessert:2023vyl} will search 
    for mono-chromatic $X$ ray emission from the radiative decay of $N_1$ dark matter in the Milky Way and nearby galaxies.
    A positive measurement will be useful as another input to discriminate between the various dark matter dilution scenarios and neutrino 
    mass generation mechanisms in the LRSM, and map out the favored parameter space.
\end{itemize}

On the theory side, a follow-up exercise will be to calculate the evolution of primordial matter 
density perturbations by taking into account the non-linear terms in the Boltzmann equations, which start to 
become non-negligible (bringing in corrections of percent level or higher) for wave numbers 
$k \gtrsim 0.2\, h/{\rm Mpc}$ and observations made at redshift $z\approx 0$~\cite{Lesgourgues:2011rh, 2012PhRvD..85l3524J, Carrasco:2013mua}. 
Because the hot dark matter component from dilutor decay acts to suppress the perturbations, we expect the non-linear effect to 
be smaller than the case of regular cold/warm dark matter. 
Nonetheless, a more careful analysis is warranted and can serve as a useful tool for discovering this piece of new physics in the 
upcoming precision cosmology era.

\section*{Acknowledgements}
We thank Weiyi Deng, Diego Redigolo, Filippo Sala and Katelin Schutz for useful discussions and correspondence.
MN is supported by the Slovenian Research Agency under the research core funding No. P1-0035 and in part 
by the research grants J1-3013, N1-0253 and J1-4389. 
YZ is supported by a Subatomic Physics Discovery Grant (individual) from the Natural Sciences and Engineering 
Research Council of Canada, and by the Canada First Research Excellence Fund through the Arthur B. McDonald Canadian Astroparticle 
Physics Research Institute.

\appendix
%
%
\section{\texorpdfstring{Radiative $N_1$ decay via $W-W_R$ mixing}{Radiative N1 decay via WL-WR mixing}}
\label{appdx1}

\begin{figure}[ht]
  \begin{center}
    \includegraphics[width=0.816\textwidth]{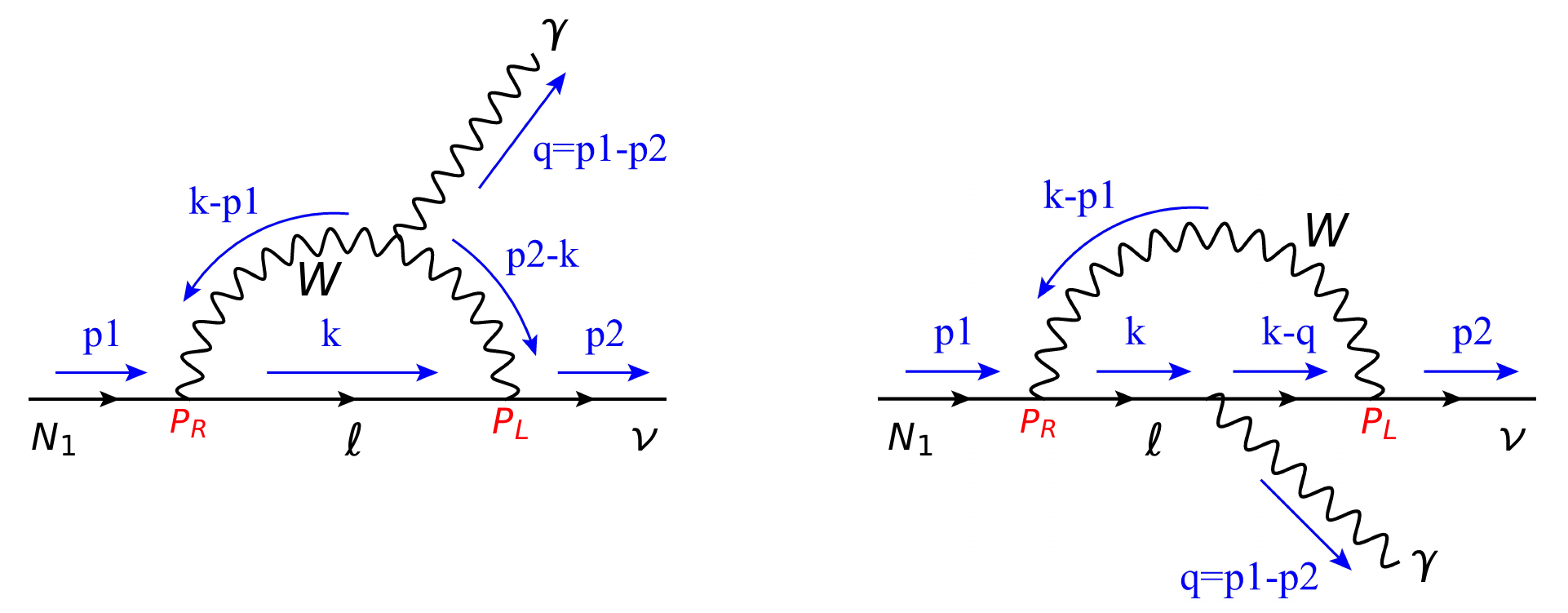}\vspace{-0.7cm}
  \end{center}
  \caption{
  Feynman diagrams for $N_1$ radiative decay via $W-W_R$ mixing in the LRSM. The $W$-$N_1$-$\ell$ vertex labelled by $\mathbbm{P}_R$ is induced by the $W$-$W_R$ mixing and the corresponding Feynman rule is obtained from the second term in Eq.~\eqref{eq:A1}.}\label{fig:N1decayW}
\end{figure}

The radiative decay of dark matter to a monochromatic photon $N_1 \to \nu \gamma$ gives a very stringent constraint 
on its couplings from the X-ray spectra measurements, shown in Fig.~\ref{fig:type2}.
Here we provide some details on the calculation of the rate in Eq.~\eqref{eq:N_1radiativexiLR}, which
has new sources within the LRSM.
There are two possible contributions, one from the Dirac mixing of $N_1$ with $\nu$, which is well known and
the same as for the sterile neutrinos.
In the presence of gauge boson mixing $\xi_{\rm LR}$, another amplitude is present, coming from the SM-like $W$
having a coupling to the right-handed charged current. 
In the small $\xi_{\rm LR}$ limit we have
\begin{equation}\label{eq:A1}
  \mathcal{L}_{\text{CC}} \simeq \frac{g}{\sqrt{2}} \left[\bar \nu_\ell \gamma^\mu \mathbbm{P}_L \ell + 
  \xi_{\rm LR} \left( V^{R\dagger}_{\rm PMNS} \right)_{i\ell} \bar N_i \gamma^\mu \mathbbm{P}_R \ell 
  \right] W_\mu^+ + {\rm h.c.} \, ,
\end{equation}
where $V^{R\dagger}_{\rm PMNS}$ is the right-handed PMNS matrix introduced in Eq.~\eqref{eq:RHCurrent}.
With this coupling turned on, there are two new diagrams contributing to radiative $N_1$ decay, and their 
topologies are shown in Fig.~\ref{fig:N1decayW}.

The $N_1\to \nu \gamma$ decay always occurs via the  dimension-5 effective operator
\begin{equation}
  \mathcal{L}_{\rm eff} = C \, \bar \nu \sigma_{\mu\nu} \mathbbm{P}_R N_1 F^{\mu\nu} + {\rm h.c.} \, ,
\end{equation}
where $C$ is the Wilson coefficient, to be determined next, and $\sigma_{\mu\nu} \equiv \frac{i}{2}[\gamma^\mu, \gamma^\nu]$.
The chiral projection operator in front of the $N_1$ field must be $\mathbbm{P}_R = (1+\gamma_5)/2$.
The corresponding decay amplitude for $N_1(p_1)\to \nu(p_2) \gamma (q)$ is
\begin{equation} \label{eq:A3}
  i \mathcal{M} = - i C \bar u_\nu(p_2,s_2) (\cancel{q} \cancel{\varepsilon}^* - \cancel{\varepsilon}^*\cancel{q}) \mathbbm{P}_R u_N(p_1, s_1)  \ .
\end{equation}
where $\varepsilon^*_\mu$ is the photon polarization vector and $q\cdot \varepsilon^*(q) = 0$ for an 
on-shell transverse photon.
The partial decay rate of $N_1\to \nu \gamma$ is 
\begin{equation}
  \Gamma_{N_1\to \nu \gamma} = \frac{1}{4\pi} \left| C \right|^2 m_{N_1}^3 \, .
\end{equation}
For a Majorana $N_1$, it can also decay into $\bar\nu \gamma$ with the same partial rate.

With the momentum assignments shown in Fig.~\ref{fig:N1decayW}, the first diagram has an amplitude
\begin{equation} \label{eq:Ntonugamma_Amp1}
\begin{split}
  i \mathcal{M}_1 &= - \frac{e g^2}{2} \xi_{\rm LR} \sum_\ell (V^{R\dagger}_{\rm PMNS})_{1\ell} m_\ell 
  \int\frac{\text{d}^4k}{(2\pi)^4} \frac{1}{(k^2-m_\ell^2)[(k-p_1)^2 - M_W^2][(p_2-k)^2 - M_W^2]} 
  \\
  & \times \left\{ \left( 2 k - p_1 - p_2 \right) \cdot \varepsilon^* \bar u_\nu(p_2, s_2) \gamma^\mu \gamma_\mu 
  \mathbbm{P}_R u_N(p_1, s_1) \rule{0mm}{4mm} \right. 
  \\
  & + \bar u_\nu(p_2, s_2) \cancel{\varepsilon}^* (\cancel{p}_2 - \cancel{k} - \cancel{q}) 
  \mathbbm{P}_R u_N(p_1, s_1) \left. + 
  \bar u_\nu(p_2, s_2) (\cancel{q} - \cancel{k} + \cancel{p}_1) \cancel{\varepsilon}^*\mathbbm{P}_R u_N(p_1, s_1) 
  \rule{0mm}{4mm}\right\} \, ,
\end{split}
\end{equation}
where we have dropped terms that are suppressed by additional powers of $1/M_W$.
For the $\gamma$ matrices between the fermion spinors, we are interested in the structure 
$\cancel{q} \cancel{\varepsilon}^* - \cancel{\varepsilon}^*\cancel{q}$, as in Eq.~\eqref{eq:A3}. 
This immediately implies that the first term in $\{\}$ does not contribute.
In addition, because the external fermions already have the correct chirality, we can drop the chirality-flipping terms (upon equation of motion) such as $\cancel{p}_1$ acting on $u_N(p_1, s_1)$ and $\cancel{p}_2$ acting on $\bar u_\nu(p_2, s_2)$. 
This allows us to reduce the $\{\}$ bracket in Eq.~\eqref{eq:Ntonugamma_Amp1} into
\begin{equation}
\begin{split}
  \left\{ \rule{0mm}{4mm}\right\} &\to \bar u(p_2, s_2) \left( 
  \cancel{\varepsilon}^* \left( - \cancel{k} - 2\cancel{q} \right) + 
  \left( 2 \cancel{q} - \cancel{k} \right) \cancel{\varepsilon}^* \right) \mathbbm{P}_R u(p_1, s_1)
  \rule{0mm}{4mm} \, .
\end{split}
\end{equation}
After completing the $k$ integral, the remaining relevant term is
\begin{equation}\label{eq:A7}
\begin{split}
  i \mathcal{M}_1 = i \frac{e g^2}{16 \pi^2} \xi_{\rm LR} \sum_\ell (V^{R\dagger}_{\rm PMNS})_{1\ell} 
  \frac{m_\ell}{M_W^2}
  \bar u(p_2, s_2) \left(\cancel{q} \cancel{\varepsilon}^* - \cancel{\varepsilon}^*\cancel{q} \right) 
  \mathbbm{P}_R u(p_1, s_1) \, .
\end{split}
\end{equation}

The second diagram of Fig.~\ref{fig:N1decayW} has an amplitude
\begin{equation}
\begin{split}
  i \mathcal{M}_2 &= - \frac{e g^2}{2} \xi_{\rm LR} \sum_\ell (V^{R\dagger}_{\rm PMNS})_{1\ell} 
  m_\ell \int\frac{\text{d}^4 k}{(2 \pi)^4} \frac{1}{(k^2-m_\ell^2)[(k-q)^2 - m_\ell^2][(k-p_1)^2 - M_W^2]} 
  \\
  & \quad\times \left\{ \bar u_\nu(p_2, s_2) \gamma^\mu \cancel{\varepsilon}^* \cancel{k} \gamma_\mu 
  \mathbbm{P}_R u_N(p_1, s_1) + 
  \bar u_\nu(p_2, s_2) \gamma^\mu (\cancel{k} -\cancel{q}) \cancel{\varepsilon}^* \gamma_\mu 
  \mathbbm{P}_R u_N(p_1, s_1) \rule{0mm}{4mm}\right\} \, .
\end{split}
\end{equation}
Using the identity $\gamma^\mu \gamma^\beta \gamma^\rho \gamma_\mu = 4 g^{\beta\rho}$, all the $\gamma$ 
matrices between $\bar u_\nu(p_2, s_2)$ and $u_N(p_1, s_1) $ are gone. 
Thus, we conclude that this diagram dose not contribute to the $N_1 \to \nu \gamma$ decay.

Comparing Eqs.~\eqref{eq:A3} and \eqref{eq:A7}, we get
\begin{equation}
  C = - \frac{e G_F \xi_{\rm LR} \sum_\ell (V^{R\dagger}_{\rm PMNS})_{1\ell} m_\ell}{2\sqrt2 \pi^2} \, .
\end{equation}
The corresponding $N_1$ radiative decay rate is
\begin{equation}
  \Gamma_{N_1\to \nu\gamma} = \frac{\alpha \xi_\text{LR}^2}{8\pi^4} G_F^2 m_{N_1}^3 
  \sum_\ell \left| \left( V^R_{\rm PMNS} \right)_{\ell1} \right|^2 m_\ell^2 \, .
\end{equation}
This is how we get Eq.~\eqref{eq:N_1radiativexiLR} in the main text.

\bibliographystyle{apsrev4-2}
\bibliography{anatomy-wdm-lr}

\end{document}